\documentclass[aps,prc,preprint,groupedaddress,nofootinbib,showpacs]{revtex4-1}

\usepackage{amsmath}
\usepackage{amssymb}
\usepackage{graphicx}
\usepackage{dcolumn}
\usepackage{bm}
\usepackage{braket}
\usepackage{hyperref}

\begin{document}


\title{Equation of state for neutron stars in SU(3) flavor symmetry}


\author{Tsuyoshi Miyatsu}
\email[]{tmiyatsu@ssu.ac.kr}
\affiliation{Department of Physics, Soongsil University, Seoul 156-743, Korea}

\author{Myung-Ki Cheoun}
\email[]{cheoun@ssu.ac.kr}
\affiliation{Department of Physics, Soongsil University, Seoul 156-743, Korea}

\author{Koichi Saito}
\email[]{koichi.saito@rs.tus.ac.jp}
\affiliation{Department of Physics, Faculty of Science and Technology, Tokyo University of Science (TUS), Noda 278-8510, Japan,}
\affiliation{J-PARC Branch, KEK Theory Center, Institute of Particle and Nuclear Studies, KEK, Tokai 319-1106, Japan}

\date{\today}

\begin{abstract}
Using several relativistic mean field models (such as GM1, GM3, NL3, TM1, FSUGold and IU-FSU) as well as the quark-meson coupling model, 
we calculate the particle fractions, the equation of state, the maximum mass and radius of a neutron star within relativistic Hartree approximation.  
In determining the couplings of the isoscalar, vector mesons to the octet baryons, we examine the extension of SU(6) spin-flavor symmetry to SU(3) flavor symmetry. 
Furthermore, we consider the strange ($\sigma^{\ast}$ and $\phi$) mesons, and study how they affect the equation of state.  
We find that the equation of state in SU(3) symmetry can sustain a neutron star with mass of $(1.8 \sim 2.1) M_{\odot}$ even if hyperons exist inside the core. 
In addition, the strange vector ($\phi$) meson and the variation of baryon structure in matter also play important roles in supporting a massive neutron star.
\end{abstract}

\pacs{26.60.Kp, 97.60.Jd, 21.80.+a, 21.30.Fe}
\keywords{}

\maketitle

\section{Introduction}
\label{sec:Introduction}

Neutron stars, which comprise hadrons and leptons as remnants of supernovae explosions, may be believed to be cosmological laboratories for dense nuclear matter. 
However, their detailed properties, for instance, the mass, radius and particle fractions in the core of a neutron star, are not fully understood yet, 
since the pioneering paper by Baade and Zwicky~\cite{PhysRev.46.76.2} and the first discovery of a neutron star by Hewish and Okoye~\cite{Nature.207.59H}.
Because the observed mass and/or radius of a neutron star can provide strong constraints on the equation of state (EoS) of dense nuclear matter, 
many theoretical discussions have been focused on the EoS to understand the structure of dense matter. 

The typical mass of neutron stars is known to be around $1.4 M_{\odot}$ ($M_{\odot}$: the solar mass)~\cite{Lattimer:2006xb}.
The most famous, precisely observed pulsar is the binary pulsar PSR B1913+16 (the Hulse-Taylor pulsar) 
with the mass of $1.4398\pm0.0002 M_{\odot}$~\cite{Taylor:1989sw,Weisberg:2010zz}.
However, a few neutron stars whose masses are much heavier than $1.4 M_{\odot}$ have recently been observed. 
For example, Shapiro delay measurements have indicated that the binary millisecond pulsar PSR J1614-2230 has the mass of $1.97\pm0.04 M_{\odot}$~\cite{Demorest:2010bx}. 
Then, such heavy neutron stars have attracted a lot of interest not only in astrophysics but also in nuclear physics, because of the possibility of exotic degrees of freedom, 
such as quarks, gluons and/or some unusual condensations of boson-like matter, in the core.

Recently, many people have often used relativistic mean-field (RMF) models (or relativistic Hartree models) including hyperons ($Y$) to calculate the EoS for a neutron star. 
However, it is quite difficult to explain the heavy neutron stars by such EoSs with the meson-baryon coupling constants based on SU(6) (quark model) symmetry, 
because the degrees of freedom of hyperons make the EoS very soft, 
and thus the possible maximum mass of a neutron star is considerably reduced~\cite{Glendenning:1991es,SchaffnerBielich:2008kb}.  

In Refs.~\cite{Miyatsu:2011bc,Katayama:2012ge}, 
the properties of a neutron star have been studied in detail within relativistic Hartree-Fock (RHF) approximation. 
In those calculations, we have considered not only the tensor couplings of vector mesons to the octet baryons and 
the form factors at interaction vertices but also the change of the internal (quark) structure of baryons in dense matter.  
The RHF calculations have performed in two ways: one with the coupling constants determined by SU(6) symmetry, 
the other with the coupling constants based on SU(3) (flavor) symmetry (see also Ref.~\cite{Weissenborn:2011ut}).  
Then, we have found that the baryon composition of the core matter in SU(3) symmetry is completely different from that in SU(6) symmetry.  
In SU(6) symmetry, all octet baryons appear in the density region below $\sim 1.2$ fm$^{-3}$, while, in the SU(3) calculation, only the $\Xi^{-}$ hyperon is produced. 
Furthermore, the medium modification of the internal baryon structure hardens the EoS for the core.  
Taking all those effects into account, we have obtained the maximum mass of a neutron star which is consistent with PSR J1614-2230.  
Therefore, it is very vital to consider the extension from SU(6) symmetry to SU(3) symmetry and  
the effect of the internal baryon-structure variation in nuclear matter. 

Now it is interesting to construct the EoS based on SU(3) symmetry in RMF approximation, 
and to see how the symmetry extension affects the EoS, 
because the RMF calculation is practically much simpler than the RHF one, and many people have thus proposed many useful RMF models, 
some of which are accurately calibrated by various experimental data not only on infinite nuclear matter but also on finite nuclei.  
In this paper, we extend several popular RMF models, such as the GM1, GM3, NL3, TM1, FSUGold and IU-FSU models, 
and study the properties of nuclear matter and the mass-radius relations of neutron stars using the isoscalar, vector-meson couplings to the octet baryons in SU(3) symmetry.  
We then compare the results in SU(3) symmetry with those in the (usual) SU(6) calculations. 

In addition, we propose RMF models including the variation of baryon structure in a dense medium.  
In such models, we also use the coupling constants determined in SU(3) symmetry, and compare the results with those calculated in SU(6) symmetry.  
To take the variation of the in-medium baryon structure into account, 
we use the quark-meson coupling (QMC)~\cite{Guichon:1987jp, Saito:1994ki} and the chiral quark-meson coupling (CQMC)~\cite{Nagai:2008ai} models.  
It is well recognized that the constituent quark mass in a hadron is generally given by the quark condensate, $\langle {\bar q} q \rangle$.  
The quark mass (or $\langle {\bar q} q \rangle$) in nuclear matter may then be reduced from the value in vacuum, 
because of the condensed scalar ($\sigma$) field depending on the nuclear density, namely the Lorentz-scalar, attractive interaction in nuclear matter.  
The decrease of the quark mass leads to the variation of baryon internal structure at the quark level.  
Such an effect is considered self-consistently in the QMC model.  

The CQMC model is an extended version of the QMC model, in which the quark-quark hyperfine interaction caused by the one-gluon exchange is included.  
In addition, the pion-exchange interaction based on chiral symmetry is also considered.  
The hyperfine interaction plays an impotent role in the baryon spectra in matter~\cite{Nagai:2008ai,Saito:2010zw}. 
The QMC and CQMC models have been successfully applied in studying the properties of hadrons in nuclear matter~\cite{Saito:1996yb}, 
finite nuclei~\cite{Saito:1996sf,Guichon:1996}, 
hypernuclei~\cite{Tsushima:1997cu, Miyatsu:2010zz} and neutron stars~\cite{Miyatsu:2011bc,Katayama:2012ge,Miyatsu:2012xh}. 
(For a review, see Ref.~\cite{Saito:2007rv}.) 

Using those models, we calculate the particle fractions, the meson fields and the EoS inside the core.  
Furthermore, we estimate the maximum mass and radius of a neutron star by solving the the Tolman-Oppenheimer-Volkoff (TOV) equation~\cite{Tolman:1934za,Oppenheimer:1939ne}. 
In the present calculations, we also study the role of the strange mesons ($\sigma^{\ast}$ and $\phi$) in the EoS.  
In SU(3) symmetry, we then find that the models except for GM3, FSUGold and IU-FSU can explain the mass of PSR J1614-2230.  
In the GM3, FSUGold and IU-FSU models, although the maximum mass cannot reach $1.97\pm0.04 M_{\odot}$, the calculated mass is not far from that value.  
Therefore, the extension from SU(6) to SU(3) symmetry is very vital for sustaining a heavy neutron star. 
In addition, the strange vector-meson ($\phi$) and the medium variation of baryon structure also help prevent the collapse of a neutron star. 

In RMF models, various types of nonlinear potentials with respect to the meson fields are usually involved, 
and they are very significant to reproduce the saturation condition for symmetric nuclear matter and the properties of finite nuclei.  
Among them, especially the $c_{3} \omega^{4}$ term hardens the EoS at high density, and thus enhances a neutron-star mass.  
Furthermore, the nonlinear isoscalar-isovector coupling, $\Lambda_{\omega \rho} \omega^{2} \rho^{2}$, 
which is involved only in the FSUGold and IU-FSU models, plays a unique role in the particle fractions in the core. 
If the $\sigma$-$\Sigma$ and $\sigma^{\ast}$-$\Sigma$ coupling constants are determined so as to fit the (repulsive) mean-field potential for the $\Sigma$ in nuclear matter, 
the $\Sigma$ hyperon usually tends to be excluded in the core of a neutron star.  
However, in the FSUGold and IU-FSU models, the $\Sigma^{-}$ as well as the $\Lambda$ and $\Xi^{-}$ can emerge with a considerable fraction even at rather low density, 
which may be caused by the $\Lambda_{\omega \rho} \omega^{2} \rho^{2}$ interaction. 

This paper is organized as follows.  
In Section~\ref{sec:extended-RMF-model}, a brief review for RMF models based on Quantum Hadrodynamics (QHD)~\cite{Serot:1984ey} is presented.  
The usual RMF models, the QMC and CQMC models are then unified through the scalar polarizability. 
In Section~\ref{sec:SU3-symmetry}, the SU(3) extension in the coupling constants of the isoscalar, vector mesons is explained. 
The parameters in various models are determined in Section~\ref{sec:models}.  
Numerical results and discussions are addressed in Section~\ref{sec:results-and-discussions}. Finally, we give a summary in Section~\ref{sec:summary}.  

\section{Relativistic mean-field models}
\label{sec:extended-RMF-model}

For describing the properties of the core of a neutron star,
we extend the usual Lagrangian density in RMF approximation to include not only the $\sigma$, $\omega$ and $\vec{\rho\,}$ mesons but also the strange mesons, 
namely the isoscalar, Lorentz scalar ($\sigma^{\ast}$) and vector ($\phi$) mesons. 
The $\sigma^{\ast}$ and $\phi$ mesons are predominantly comprised of $\bar{s}s$ quarks. 
Because the charge neutrality and $\beta$ equilibrium conditions are imposed in the core, the leptons must be introduced as well.
The Lagrangian density is thus chosen to be
\begin{eqnarray}
	\mathcal{L}
	&=& \sum_{B}\bar{\psi}_{B}\left[
    i\gamma_{\mu}\partial^{\mu}
	- M_{B}^{\ast}\left(\sigma,\sigma^{\ast}\right)
	- g_{\omega B}\gamma_{\mu}\omega^{\mu}
	- g_{\phi B}\gamma_{\mu}\phi^{\mu}
	- g_{\rho B}\gamma_{\mu}\vec{\rho}^{\,\mu}\cdot\vec{I}_{B} \right]\psi_{B}
	\nonumber \\
	&+& \frac{1}{2} \left( \partial_{\mu}\sigma\partial^{\mu}\sigma-m_{\sigma}^{2}\sigma^{2} \right)
	+ \frac{1}{2} \left( \partial_{\mu}\sigma^{\ast}\partial^{\mu}\sigma^{\ast}
	- m_{\sigma^{\ast}}^{2}\sigma^{\ast2} \right)
	\nonumber \\
	&+& \frac{1}{2}m_{\omega}^{2}\omega_{\mu}\omega^{\mu} - \frac{1}{4}W_{\mu\nu}W^{\mu\nu}
	+ \frac{1}{2}m_{\phi}^{2}\phi_{\mu}\phi^{\mu} - \frac{1}{4}P_{\mu\nu}P^{\mu\nu}
	+ \frac{1}{2}m_{\rho}^{2}\vec{\rho}_{\mu}\cdot\vec{\rho}^{\,\mu}
	- \frac{1}{4}\vec{R}_{\mu\nu}\cdot\vec{R}^{\mu\nu}
	\nonumber \\
	&-& U_{NL}(\sigma,\omega^{\mu},\vec{\rho}^{\,\mu})
	+ \sum_{\ell}\bar{\psi}_{\ell}\left[i\gamma_{\mu}\partial^{\mu}-m_{\ell}\right]\psi_{\ell} \ ,
  \label{eq:total-Lagrangian-density}
\end{eqnarray}
where
\begin{equation}
	W_{\mu\nu}      = \partial_{\mu}\omega_{\nu} - \partial_{\nu}\omega_{\mu} \ , \ \ \ 
	P_{\mu\nu}       = \partial_{\mu}\phi_{\nu} - \partial_{\nu}\phi_{\mu} \ , \ \ \ 
	\vec{R}_{\mu\nu} = \partial_{\mu}\vec{\rho}_{\nu} - \partial_{\nu}\vec{\rho}_{\mu} \ , 
	\label{eq:covariant-derivative}
\end{equation}
with $\psi_{B (\ell)}$ the baryon (lepton) field,
$\vec{I}_B$ the isospin matrix for baryon $B$ and $m_{\ell}$ the lepton mass.
The sum $B$ runs over the octet baryons, $N$ (proton and neutron), $\Lambda$, $\Sigma^{+,0,-}$ and $\Xi^{0,-}$, and the sum $\ell$ is for the leptons, $e^{-}$ and $\mu^{-}$.
The $\omega$-, $\phi$- and $\rho$-$B$ coupling constants are respectively denoted by $g_{\omega B}$, $g_{\phi B}$ and $g_{\rho B}$.  
In Eq.~(\ref{eq:total-Lagrangian-density}), $U_{NL}$ is a nonlinear potential, which is explained below. 

When the baryons are treated as point-like objects (as in QHD), the effective baryon mass, $M_{B}^{\ast}$, in matter is simply expressed as
\begin{equation}
	M_{B}^{\ast}\left(\sigma,\sigma^{\ast}\right)
	= M_{B} - g_{\sigma B}\sigma - g_{\sigma^{\ast}B}\sigma^{\ast} \ ,
	\label{eq:effective-mass-QHD}
\end{equation}
where $M_{B}$ is the mass in vacuum,
and $g_{\sigma B}$ and $g_{\sigma^{\ast}B}$ are the $\sigma$- and $\sigma^{\ast}$-$B$ coupling constants, respectively. 
We hereafter call the model in which the baryons are structureless the QHD-type model. 

In contrast, in the QMC and CQMC models, 
the coupling constants, $g_{\sigma B}$ and $g_{\sigma^{\ast}B}$, depend on the $\sigma$ and $\sigma^{\ast}$ fields, 
which reflects the variation of internal baryon structure in matter~\cite{Saito:1994ki,Saito:2007rv}.  
Such dependences are caused by the attractive interactions due to the $\sigma$ and $\sigma^{\ast}$ exchanges. 
Thus, the in-medium baryon mass can be written as~\cite{Miyatsu:2012xh}
\begin{equation}
	M_{B}^{\ast}\left(\sigma,\sigma^{\ast}\right)
	= M_{B} - g_{\sigma B}(\sigma)\sigma - g_{\sigma^{\ast}B}(\sigma^{\ast})\sigma^{\ast} \ ,
	\label{eq:effective-mass-QMC}
\end{equation}
with the following, simple parameterizations~\cite{Guichon:1996,Saito:1996yb,Miyatsu:2010zz,Miyatsu:2011bc,Katayama:2012ge} 
\begin{eqnarray}
	g_{\sigma B}(\sigma)
	&=& g_{\sigma B}b_{B}\left[1-\frac{a_{B}}{2}\left(g_{\sigma N}\sigma\right)\right] \ ,
	\label{eq:cc-sigma} \\
	g_{\sigma^{\ast}B}(\sigma^{\ast})
	&=& g_{\sigma^{\ast}B}b_{B}^{\prime}
	\left[1-\frac{a_{B}^{\prime}}{2}\left(g_{\sigma^{\ast}\Lambda}\sigma^{\ast}\right)\right] \ ,
	\label{eq:cc-sigma-star}
\end{eqnarray}
where $g_{\sigma N}$ and $g_{\sigma^{\ast}\Lambda}$ are respectively
the $\sigma$-$N$ and $\sigma^{\ast}$-$\Lambda$ coupling constants at zero density.
Here, we introduce four parameters, $a_{B}$, $b_{B}$, $a_{B}^{\prime}$ and $b_{B}^{\prime}$,
for describing the mass, and their values are tabulated in Table~\ref{tab:QMC-parameter}.
\begin{table}
\caption{\label{tab:QMC-parameter}
Values of $a_{B}$, $b_{B}$, $a_{B}^{\prime}$ and $b_{B}^{\prime}$ for the octet baryons in the QMC or CQMC model.
We assume that the scalar, strange ($\sigma^{\ast}$) meson does not couple to the nucleon. 
}
\begin{ruledtabular}
\begin{tabular}{lcccccccc}
\         & \multicolumn{4}{c}{QMC}                                           & \multicolumn{4}{c}{CQMC}                                          \\
$B$       & $a_{B}$~(fm) & $b_{B}$ & $a_{B}^{\prime}$~(fm) & $b_{B}^{\prime}$ & $a_{B}$~(fm) & $b_{B}$ & $a_{B}^{\prime}$~(fm) & $b_{B}^{\prime}$ \\
\hline
$N$       & 0.179        & 1.00    & ---                   & ---              & 0.118        & 1.04    & ---                   & ---              \\
$\Lambda$ & 0.172        & 1.00    & 0.220                 & 1.00             & 0.122        & 1.09    & 0.290                 & 1.00             \\
$\Sigma$  & 0.177        & 1.00    & 0.223                 & 1.00             & 0.184        & 1.02    & 0.277                 & 1.15             \\
$\Xi$     & 0.166        & 1.00    & 0.215                 & 1.00             & 0.181        & 1.15    & 0.292                 & 1.04             \\
\end{tabular}
\end{ruledtabular}
\end{table}
The effect of the variation of baryon structure at the quark level can be described with 
the parameters $a_{B}$ and $a_{B}^{\prime}$.  In addition, in the CQMC model, the extra parameters, $b_{B}$ and $b_{B}^{\prime}$, 
are necessary to express the effect of hyperfine interaction between two quarks~\cite{Nagai:2008ai,Saito:2010zw,Miyatsu:2010zz}.
If we set $a_{B}=0$ and $b_{B}=1$, $g_{\sigma B}(\sigma)$ becomes identical to the $\sigma$-$B$ coupling constant in QHD.
This is also true of the coupling $g_{\sigma^{\ast}B}(\sigma^{\ast})$.

In the QHD-type model, we add the following nonlinear (NL) potential to the Lagrangian density
\begin{equation}
	U_{NL}(\sigma,\omega^{\mu},\vec{\rho}^{\,\mu})
	= \frac{1}{3}g_{2}\sigma^{3}
	+ \frac{1}{4}g_{3}\sigma^{4}
	- \frac{1}{4}c_{3}\left(\omega_{\mu}\omega^{\mu}\right)^{2}
	- \Lambda_{\omega\rho}
	\left( \omega_{\mu}\omega^{\mu}\right)\left(\vec{\rho}_{\mu}\cdot\vec{\rho}^{\,\mu} \right) \ ,
	\label{eq:Lagrangian-NL}
\end{equation}
so as to reproduce the measured properties of nuclear matter and finite nuclei, for example, 
the incompressibility of nuclear matter, $K_v$, the density dependence of symmetry energy, $a_4$, etc.  
Here, the potential involves four coupling constants, $g_2$, $g_3$, $c_{3}$ and $\Lambda_{\omega\rho} (\equiv \Lambda_v g_{\rho N}^2 g_{\omega N}^2)$. 

In RMF approximation, the meson fields are replaced by the constant mean-field values:
$\bar{\sigma}$, $\bar{\omega}$, $\bar{\sigma}^{\ast}$, $\bar{\phi}$ and $\bar{\rho}$ (the $\rho^{0}$ field).
The equations of motion for the meson fields in uniform matter are thus given by
\begin{eqnarray}
	m_{\sigma}^{2}\bar{\sigma} + g_{2}\bar{\sigma}^{2} + g_{3}\bar{\sigma}^{3}
	&=& \sum_{B} g_{\sigma B} C_{B}(\bar{\sigma}) \rho_{B}^s \ ,
	\label{eq:EOM-sigma} \\
	m_{\sigma^{\ast}}^{2}\bar{\sigma}^{\ast}
	&=& \sum_{B} g_{\sigma^{\ast} B} C_{B}^{\prime}(\bar{\sigma}^{\ast}) \rho_{B}^s \ ,
	\label{eq:EOM-sigma-star} \\
	\left(m_{\omega}^{2}+2\Lambda_{\omega\rho}\bar{\rho}^{2}\right)\bar{\omega} + c_{3}\bar{\omega}^{3}
	&=& \sum_{B} g_{\omega B} \rho_{B} \ ,
	\label{eq:EOM-omega} \\
	m_{\phi}^{2}\bar{\phi}
	&=& \sum_{B} g_{\phi B} \rho_{B} \ ,
	\label{eq:EOM-phi} \\
	\left(m_{\rho}^{2}+2\Lambda_{\omega\rho}\bar{\omega}^{2}\right)\bar{\rho}
	&=& \sum_{B} g_{\rho B} (\vec{I}_{B})_{3} \rho_{B} \ ,
	\label{eq:EOM-rho}
\end{eqnarray}
where the scalar density, $\rho_{B}^{s}$, and the baryon density, $\rho_{B}$, read 
\begin{eqnarray}
	\rho_{B}^{s}
	&=& \frac{1}{\pi^{2}}\int_{0}^{k_{F_{B}}}dk~k^{2}
	\frac{M_{B}^{\ast}(\bar{\sigma},\bar{\sigma}^{\ast})}
	{\left[k^{2}+M_{B}^{\ast2}(\bar{\sigma},\bar{\sigma}^{\ast})\right]^{1/2}} \ ,
	\label{eq:baryon-scalar-density} \\
	\rho_{B}
	&=& \frac{1}{\pi^{2}}\int_{0}^{k_{F_{B}}}dk~k^{2}
	= \frac{k^{3}_{F_{B}}}{3\pi^{2}} \ ,
	\label{eq:baryon-density}
\end{eqnarray}
with $k_{F_B}$ being the Fermi momentum for baryon $B$.  

In Eqs.~(\ref{eq:EOM-sigma}) and (\ref{eq:EOM-sigma-star}), 
$C_{B}$ and $C_{B}^{\prime}$ are respectively  the scalar polarizabilities (or the scalar-density ratios) at the $\sigma$-$B$ and $\sigma^{\ast}$-$B$ interactions.  
Here, the scalar polarizabilities is defined by the ratio of the scalar density of a confined quark field at finite density to that in vacuum.  
In the QMC or CQMC model, they can be expressed by the following parameterizations~\cite{Miyatsu:2011bc,Katayama:2012ge,Miyatsu:2012xh,Tsushima:1997cu}:
\begin{eqnarray}
	C_{B}(\bar{\sigma})
	&=& b_{B} \left[ 1 - a_{B}\left(g_{\sigma N} \bar{\sigma} \right) \right] \ ,
	\label{eq:scalar-density-ratio-sigma} \\
	C_{B}^{\prime}(\bar{\sigma}^{\ast})
	&=& b_{B}^{\prime} \left[ 1 - a_{B}^{\prime}\left(g_{\sigma^{\ast}\Lambda} \bar{\sigma}^{\ast} \right) \right] \ ,
	\label{eq:scalar-density-ratio-sigma-star}
\end{eqnarray}
where the parameters $a_{B}$, $b_{B}$, $a_{B}^{\prime}$ and $b_{B}^{\prime}$ take the same values as 
in Eqs.~(\ref{eq:cc-sigma}) and (\ref{eq:cc-sigma-star}) (see also Table~\ref{tab:QMC-parameter}). 
In contrast, they become unity in the QHD-type model (recall $a_{B}=a_{B}^\prime=0$ and $b_{B}=b_{B}^\prime=1$). 

The total energy density, $\epsilon$, and pressure, $P$, in the core then read 
\begin{eqnarray}
	\epsilon
	&=& \sum_{B}\frac{1}{\pi^{2}}\int_{0}^{k_{F_{B}}}dk~k^{2}
		\left[k^{2}+M_{B}^{\ast2}(\bar{\sigma},\bar{\sigma}^{\ast})\right]^{1/2}
	\nonumber \\
	&+& \frac{1}{2}m_{\sigma}^{2}\bar{\sigma}^{2}
	+   \frac{1}{3}g_{2}\bar{\sigma}^{3}
	+   \frac{1}{4}g_{3}\bar{\sigma}^{4}
	+   \frac{1}{2}m_{\sigma^{\ast}}^{2}\bar{\sigma}^{\ast2}
	\nonumber \\
	&+& \frac{1}{2}m_{\omega}^{2}\bar{\omega}^{2}
	+   \frac{3}{4}c_{3}\bar{\omega}^{4}
	+   \frac{1}{2}m_{\phi}^{2}\bar{\phi}^{2}
	+   \frac{1}{2}m_{\rho}^{2}\bar{\rho}^{2}
	+   3\Lambda_{\omega\rho} \bar{\omega}^{2} \bar{\rho}^{2}
	\nonumber \\
	&+& \sum_{\ell}\frac{1}{\pi^{2}}\int_{0}^{k_{F_{\ell}}}dk~k^{2}
		\left[k^{2}+m_{\ell}^{2}\right]^{1/2} \ ,
	\label{eq:engy-density} \\
	P
	&=& n_{B}^{2}\frac{\partial}{\partial n_{B}}\left(\frac{\epsilon}{n_{B}}\right) \ ,
	\label{eq:baryon-pressure}
\end{eqnarray}
where the total baryon density, $n_{B}$, is given by a sum of each baryon density  
\begin{equation}
	n_{B} = \sum_{B}\rho_{B} \ .
	\label{eq:total-baryon-density}
\end{equation}
In the QHD-type model, the pressure can also be expressed as 
\begin{eqnarray}
	P
	&=& \frac{1}{3} \sum_{B}\frac{1}{\pi^{2}}\int_{0}^{k_{F_{B}}}dk~\frac{k^{4}}
  		{\left[k^{2}+M_{B}^{\ast2}(\bar{\sigma},\bar{\sigma}^{\ast})\right]^{1/2}}
	\nonumber \\
	&-& \frac{1}{2}m_{\sigma}^{2}\bar{\sigma}^{2}
	-   \frac{1}{3}g_{2}\bar{\sigma}^{3}
	-   \frac{1}{4}g_{3}\bar{\sigma}^{4}
	-   \frac{1}{2}m_{\sigma^{\ast}}^{2}\bar{\sigma}^{\ast2}
	\nonumber \\
	&+& \frac{1}{2}m_{\omega}^{2}\bar{\omega}^{2}
	+   \frac{1}{4}c_{3}\bar{\omega}^{4}
	+   \frac{1}{2}m_{\phi}^{2}\bar{\phi}^{2}
	+   \frac{1}{2}m_{\rho}^{2}\bar{\rho}^{2}
	+   \Lambda_{\omega\rho} \bar{\omega}^{2} \bar{\rho}^{2}
	\nonumber \\
	&+& \frac{1}{3} \sum_{\ell}\frac{1}{\pi^{2}}\int_{0}^{k_{F_{\ell}}}dk~
		\frac{k^{4}}{\left[k^{2}+m_{\ell}^{2}\right]} \ .
	\label{eq:baryon-pressure-QHD}
\end{eqnarray}
%

\section{SU(3) symmetry in the isoscalar, vector-meson couplings}
\label{sec:SU3-symmetry}

To study the EoS and the properties of neutron stars, 
it is very interesting to extend SU(6) spin-flavor symmetry based on the quark model to the more general SU(3) flavor symmetry~\cite{Weissenborn:2011ut,Sulaksono:2012ny}.  
Restricting our interest to three quark flavors (up, down and strange), SU(3) symmetry can be regarded as a symmetry group of strong interaction.  
To consider combinations of the meson-baryon couplings, it is extremely useful to choose the SU(3)-invariant interaction Lagrangian.  
Using the matrix representations for the baryon octet, $B$, and meson nonet (singlet state, $M_{1}$, and octet state, $M_{8}$), 
the interaction Lagrangian can be written as a sum of three terms, 
namely one coming from the coupling of the meson singlet to the baryon octet ($S$ term) and the other two terms from the interaction of the meson octet and the baryons
-- one being the antisymmetric ($F$) term and the other being the symmetric ($D$) term~\cite{Weissenborn:2011ut,Rijken:2010zzb,deSwart:1963gc}:
\begin{equation}
	\mathcal{L}_{int} =
	- g_{8} \sqrt{2} \left[ \alpha {\rm Tr}\left(\left[\bar{B},M_{8}\right] B\right)
	+ (1-\alpha) {\rm Tr}\left(\left\{\bar{B},M_{8}\right\} B\right) \right]
	- g_1 \frac{1}{\sqrt{3}} {\rm Tr}\left(\bar{B}B\right){\rm Tr}\left(M_{1}\right) \ ,
	\label{eq:SU3-Lagrangian}
\end{equation}
where $g_{1}$ and $g_{8}$ are respectively the coupling constants for the meson singlet and octet states, and $\alpha$ ($0 \leq \alpha \leq 1$) is known as the $F/(F+D)$ ratio.  
For details, see the references~\cite{Rijken:2010zzb,deSwart:1963gc}.

We here focus on the isoscalar, vector-meson ($\omega$ and $\phi$) couplings to the octet baryons, 
because, as usual, the other coupling constants can be determined 
so as to reproduce the observed properties of nuclear matter and hypernuclei (as discussed in section \ref{sec:models})\footnote{
When SU(3) symmetry is applied to the {\it isovector}, vector mesons, 
the Fock term is, in fact, necessary to reproduce the observed symmetry energy~\cite{Katayama:2012ge}. }.  
The physical $\omega$ and $\phi$ mesons are described in terms of the pure singlet, $\ket{1}$, and octet, $\ket{8}$, states as 
\begin{equation}
	\omega =   \cos \theta_{v} \ket{1} + \sin \theta_{v} \ket{8} \ , \ \ \ 
	\phi   = - \sin \theta_{v} \ket{1} + \cos \theta_{v} \ket{8} \ , 
	\label{eq:mixing}
\end{equation}
with $\theta_{v}$ being the mixing angle.  

In SU(3) symmetry, all possible combinations of the couplings are then determined by four parameters:
the singlet and octet coupling constants, $g_{1}$ and $g_{8}$, the $F/(F+D)$ ratio for the vector mesons, $\alpha_{v}$, and the mixing angle, $\theta_{v}$.  
If we require the {\it universality} assumption for the (electric) $F/(F+D)$ ratio, we find $\alpha_{v}=1$~\cite{Rijken:2010zzb,Rijken:1998yy}.  
In the limit of the {\it ideal} mixing, the mixing angle is given by
\begin{equation}
	\theta_{v}^{ideal} =
	\tan^{-1} \left( \frac{1}{\sqrt{2}} \right)
	\simeq 35.26^{\circ} \ .
	\label{eq:ideal-mixing-angle}
\end{equation}
Furthermore, if the coupling ratio, $z$, is chosen to be 
\begin{equation}
	z \equiv \frac{g_8}{g_1} = \frac{1}{\sqrt{6}} \simeq 0.4082 \ ,
	\label{eq:ideal-mixing-coupling-ratio}
\end{equation}
we can obtain the usual SU(6) relations:  
\begin{eqnarray}
	\frac{1}{3} g_{\omega N} = \frac{1}{2} g_{\omega\Lambda} &=& \frac{1}{2} g_{\omega\Sigma} = g_{\omega\Xi} \ ,
	\label{eq:SU6-relation-omega} \\
	2 g_{\phi\Lambda} = 2 g_{\phi\Sigma} = g_{\phi\Xi} &=& \frac{2\sqrt{2}}{3} g_{\omega N} \ ,
	\ \ \  g_{\phi N} = 0 \ .
	\label{eq:SU6-relation-phi}
\end{eqnarray}

In the present calculation, we refer to the Nijmegen extended-soft-core (ESC) model~\cite{Rijken:2010zzb} to fix the mixing angle and $z$. 
At present, this model may be the most complete model for the baryon-baryon interaction.
It can well describe not only the $N$-$N$ but also the $Y$-$N$ and $Y$-$Y$ interactions in terms of the meson exchanges based on SU(3) symmetry.
This model has then suggested the values of $\theta_{v}$ and $z$ as 
\begin{equation}
	\theta_{v} = 37.50^{\circ} \ ,  \ \ \ 
	z = 0.1949 \ .
	\label{eq:ESC-mixing-data}
\end{equation}
We notice that the mixing angle is very close to the ideal value, while the value of $z$ is much smaller than that in SU(6) symmetry.  
It may be expected that a small value of $z$ helps enhance the coupling constants~\cite{Weissenborn:2011ut}.  
We can find the relations of the coupling constants in SU(3) symmetry as
\begin{eqnarray}
	g_{\omega\Lambda} = g_{\omega\Sigma}
	&=& \frac{1}{1+\sqrt{3}z\tan\theta_{v}} g_{\omega N} \ ,
	\hspace{0.5cm}
	g_{\omega\Xi}
	= \frac{1-\sqrt{3}z\tan\theta_{v}}{1+\sqrt{3}z\tan\theta_{v}} g_{\omega N} \ ,
	\label{eq:ESC-relations1} \\
	g_{\phi N}
	&=& \frac{\sqrt{3}z-\tan\theta_{v}}{1+\sqrt{3}z\tan\theta_{v}} g_{\omega N} \ ,
	\label{eq:ESC-relations2} \\
	g_{\phi\Lambda} = g_{\phi\Sigma}
	&=& \frac{-\tan\theta_{v}}{1+\sqrt{3}z\tan\theta_{v}} g_{\omega N} \ ,
	\hspace{0.5cm}
	g_{\phi\Xi}
	= -\frac{\sqrt{3}z+\tan\theta_{v}}{1+\sqrt{3}z\tan\theta_{v}} g_{\omega N} \ .
	\label{eq:ESC-relations3}
\end{eqnarray}
Therefore, once the value of $g_{\omega N}$ is given, 
the other coupling constants, $g_{\omega Y}$ and $g_{\phi B}$, are determined by Eqs.~(\ref{eq:ESC-relations1})-(\ref{eq:ESC-relations3}).

\section{Models}
\label{sec:models}

We examine two types of RMF models.
One is based on the QMC and CQMC models~\cite{Miyatsu:2011bc,Katayama:2012ge,Miyatsu:2012xh}, 
in which the variation of internal baryon structure in matter is taken into account.  
In these models, it is not necessary to consider any NL potential for describing the properties of nuclear matter around the saturation density, $n_B^0$. 
The other is the QHD-type models with the NL potential given in Eq.~(\ref{eq:Lagrangian-NL}). 
In fact, we adopt the parameterizations of the GM1, GM3~\cite{Glendenning:1991es}, NL3~\cite{Lalazissis:1996rd}, TM1~\cite{Sugahara:1993wz}, 
FSUGold~\cite{ToddRutel:2005zz} and IU-FSU~\cite{Fattoyev:2010mx} models. Some of those models are very popular, 
because they are accurately calibrated by using various experimental data on infinite nuclear matter and finite nuclei.  

\subsection{SU(6) symmetry}
\label{subsec:SU(6)}

In the case of QHD-type, the coupling constants, $g_{\sigma N}$, $g_{\omega N}$ and $g_{\rho N}$, 
are determined so as to reproduce the binding energy per nucleon, $w_{0}$, and symmetry energy, $a_{4}$, at $n_{B}^{0}$.  
The parameters, $g_{2}$, $g_{3}$, $c_{3}$ and $\Lambda_{\omega\rho}$, in Eq.~(\ref{eq:Lagrangian-NL}) are chosen to be the values given in the original papers.  
For the vector-meson couplings to hyperons, 
we use the SU(6) relations given in Eqs.~(\ref{eq:SU6-relation-omega}) and (\ref{eq:SU6-relation-phi}), and the following coupling relations
\begin{equation}
	g_{\rho N} = \frac{1}{2} g_{\rho\Sigma} = g_{\rho\Xi} \ , \hspace{0.5cm} g_{\rho\Lambda} = 0 \ .
	\label{eq:SU6-relation-rho}
\end{equation}

Furthermore, assuming that the $\sigma^{\ast}$ meson does not couple to a nucleon ($g_{\sigma^{\ast}N}=0$),  
the couplings of $\sigma$-$Y$ and $\sigma^{\ast}$-$Y$ may be determined as follows.  
In RMF approximation, the potential for hyperon $Y$ in symmetric nuclear matter, $U_{Y}^{(N)}$, may be calculated as 
\begin{equation}
	U_{Y}^{(N)} =
	- g_{\sigma Y} \bar{\sigma} 
	+ g_{\omega Y} \bar{\omega} \ . 
	\label{eq:potential-depth1}
\end{equation}
Thus, we can determine the coupling constants, $g_{\sigma Y}$, 
if we take the following values suggested from the experimental data of hypernuclei:  
$U_{\Lambda}^{(N)} = -28$ MeV, $U_{\Sigma}^{(N)} = +30$ MeV and $U_{\Xi}^{(N)} = -18$ MeV~\cite{Schaffner:1995th,Schaffner:1993qj,Yang:2008am}.  

In addition, if we consider the Nagara event~\cite{Nagara}, 
which may suggest that the depth of the potential between two $\Lambda$s is about $-5$ MeV, 
we may be able to fix the coupling constant, $g_{\sigma^{\ast} \Lambda}$, by assuming that $U_{\Lambda}^{(\Lambda)} \simeq -5$ MeV, 
where $U_{\Lambda}^{(\Lambda)}$ is the potential for $\Lambda$ in $\Lambda$-hyperon matter: 
\begin{equation}
	U_{\Lambda}^{(\Lambda)} =
	- g_{\sigma \Lambda} \bar{\sigma}^{(\Lambda)}
	- g_{\sigma^{\ast} \Lambda} \bar{\sigma}^{\ast(\Lambda)}
	+ g_{\omega \Lambda} \bar{\omega}^{(\Lambda)}
	+ g_{\phi \Lambda} \bar{\phi}^{(\Lambda)} \ . 
	\label{eq:potential-depth2}
\end{equation}
Here, the superscript $(Y)$ stands for a quantity in $Y$-hyperon matter.  
Furthermore, we assume the relation, $g_{\sigma^{\ast}\Sigma}=g_{\sigma^{\ast}\Lambda}$, which is presented by SU(6) symmetry, 
and determine the coupling constant, $g_{\sigma^{\ast} \Xi}$, using the relation $U_{\Xi}^{(\Xi)} \simeq 2 U_{\Lambda}^{(\Lambda)}$~\cite{Schaffner:1993qj,Yang:2008am}. 

In the QMC and CQMC models, the NL interaction is not necessary and the coupling constants can be determined by the same way as in the QHD-type model.  
We, however, notice that, in Eqs.~(\ref{eq:potential-depth1}) and (\ref{eq:potential-depth2}), 
the coupling constants for the scalar mesons should be replaced by the field-dependent ones (see Eqs.~(\ref{eq:cc-sigma}) and (\ref{eq:cc-sigma-star})).  

In Tables~\ref{tab:cc-QMC-and-CQMC}, \ref{tab:cc-non-linear-sigma} and \ref{tab:cc-non-linear-sigma-omega-rho}, 
we list the coupling constants in SU(6) symmetry and the properties of symmetric nuclear matter at $n_{B}^{0}$.  

\subsection{SU(3) symmetry}
\label{subsec:SU(3)}

As discussed in section~\ref{sec:SU3-symmetry}, because the pure singlet- and octet-states are mixed in SU(3) symmetry, 
the $\phi$ meson as well as the $\sigma$ and $\omega$ mesons contributes to the nuclear saturation properties.  
Thus, we have to readjust the coupling constants to satisfy the saturation condition, namely the binding energy per nucleon, $w_{0}$, at $n_{B}^{0}$.  
We suppose that $g_{\sigma^\ast N} = 0$ and the coupling constant, $g_{\sigma N}$, takes the same value as in SU(6) symmetry.  
The coupling constant, $g_{\rho N}$, is fixed so as to reproduce the value of symmetry energy given in the original paper.  

Firstly, we consider the QHD-type models.  
Assuming that the couplings, $g_{2}$ and $g_{3}$, in the NL potential takes the values given in the original paper, 
we can determine not only $g_{\omega N}$ and $c_{3}$ but also $g_{\phi N}$ so as to reproduce the same saturation condition as in the original paper.  
We here notice that, because $g_{\phi N}$ is related to $g_{\omega N}$ through Eq.~(\ref{eq:ESC-relations2}), $g_{\phi N}$ is not free.  
Even in the case where the quartic term of the $\omega$ field is not involved, namely $c_{3} = 0$, 
it is possible to reproduce the same saturation condition, 
because the $\phi$-meson contributions to the energy density and pressure are quadratic (see Eqs.(\ref{eq:engy-density}) and (\ref{eq:baryon-pressure-QHD})), 
and they have the same forms as in the $\omega$-meson contributions.  
For the vector-meson couplings to hyperons, we use the SU(3) relations given in Eqs.~(\ref{eq:ESC-relations1}) and (\ref{eq:ESC-relations3}), 
and the relations for the $\rho$ meson, Eq.(\ref{eq:SU6-relation-rho}).  
For the couplings of $\sigma$-$Y$ and $\sigma^{\ast}$-$Y$, we may be able to use the same procedure as in SU(6) symmetry.  

Next, in the QMC and CQMC models, we can also reproduce the same properties of nuclear matter as in SU(6) symmetry by only readjusting the coupling constant, 
$g_{\omega N}$ (thus, $g_{\phi N}$ is also varied through Eq.~(\ref{eq:ESC-relations2})).  
The other coupling constants may be determined by the same ways as in SU(6) symmetry.  

The results in SU(3) symmetry are also given in Tables~\ref{tab:cc-QMC-and-CQMC}, \ref{tab:cc-non-linear-sigma} and \ref{tab:cc-non-linear-sigma-omega-rho}.  

\begin{table}
\caption{\label{tab:cc-QMC-and-CQMC}
Coupling constants and properties of symmetric nuclear matter in the QMC and CQMC models.  
We assume that $g_{\sigma^{\ast}N}=0$ and $g_{\sigma^{\ast} \Lambda} = g_{\sigma^{\ast} \Sigma}$.  
The hadron masses are taken as follows: 
$M_{N}=939$ MeV, $M_{\Lambda}=1116$ MeV, $M_{\Sigma}=1193$ MeV, $M_{\Xi}=1318$ MeV, 
$m_{\sigma}=550$ MeV, $m_{\omega}=783$ MeV, $m_{\rho}=770$ MeV, $m_{\sigma^{\ast}}=975$ MeV and $m_{\phi}=1020$ MeV.  
The saturation condition for symmetric nuclear matter is supposed to be $w_{0}=-15.7$ MeV at $n_{B}^{0}=0.15$ fm$^{-3}$. 
The symmetry energy is taken to be $a_{4}=32.5$ MeV at $n_{B}^{0}$, and the slope parameter of the symmetry energy is denoted by $L$\footnote{
The symmetry energy, $a_4$, is defined in terms of the 2nd derivative of the total energy with respect to the difference between proton and neutron densities, 
and the slope parameter, $L$, is then given by the derivative of $a_4$ with respect to the baryon density~\cite{dutra,tsang}. 
}.  
}
\begin{ruledtabular}
\begin{tabular}{lcccc}
\                          & \multicolumn{2}{c}{QMC} & \multicolumn{2}{c}{CQMC} \\
vector sym.                & SU(6)  & SU(3)          & SU(6)  & SU(3)           \\
\hline
\multicolumn{5}{c}{--- coupling constants ---}                                  \\
$g_{\sigma N}$             & 8.28   & 8.28           & 8.50   & 8.50            \\
$g_{\omega N}$             & 8.24   & 7.98           & 9.45   & 9.14            \\
$g_{\rho N}$               & 4.38   & 4.38           & 4.29   & 4.29            \\
$g_{\phi N}$               & ---    & -2.72          & ---    & -3.12           \\
$g_{\sigma\Lambda}$        & 5.01   & 6.09           & 4.97   & 6.11            \\
$g_{\sigma\Sigma}$         & 2.45   & 3.53           & 3.24   & 4.51            \\
$g_{\sigma\Xi}$            & 2.67   & 4.83           & 2.59   & 4.84            \\
$g_{\sigma^{\ast}\Lambda}$ & 1.09   & 0.00
\footnote{
Because the $\sigma$-meson contribution in the QMC model already gives $U_{\Lambda}^{(\Lambda)}=-8$ MeV at $n_{B}^{0}$, 
the additional, attractive force due to the $\sigma^{\ast}$ meson is not required. 
}
													 & 2.62   & 1.17            \\
$g_{\sigma^{\ast}\Xi}$     & 7.53   & 5.19           & 8.46   & 5.80            \\
\multicolumn{5}{c}{--- properties of symmetric nuclear matter ---}              \\
$M_{N}^{\ast}/M_{N}$       & 0.80   & 0.80           & 0.76   & 0.76            \\
$K_{v}$ (MeV)              & 280    & 280            & 302    & 302             \\
$a_{4}$ (MeV)              & 32.5   & 32.5           & 32.5   & 32.5            \\
$L$ (MeV)                  & 88.7   & 88.7           & 90.7   & 90.7            \\
\end{tabular}
\end{ruledtabular}
\end{table}
\begin{table*}
\caption{\label{tab:cc-non-linear-sigma}
Coupling constants and properties of symmetric nuclear matter in the GM1, GM3 and NL3 models.  
The relations, $g_{\sigma^{\ast}N}=0$ and $g_{\sigma^{\ast} \Lambda} = g_{\sigma^{\ast} \Sigma}$, are assumed.  
For the NL3 model, we take $m_{\sigma}=508.194$ MeV, $m_{\omega}=782.501$ MeV and $m_{\rho}=763.000$ MeV~\cite{Lalazissis:1996rd}.  
The other masses are the same as in Table~\ref{tab:cc-QMC-and-CQMC}. 
}
\begin{ruledtabular}
\begin{tabular}{lccccccc}
\                          & \multicolumn{2}{c}{GM1} & \multicolumn{2}{c}{GM3} & \multicolumn{2}{c}{NL3} \\
vector sym.                & SU(6)  & SU(3)          & SU(6)  & SU(3)          & SU(6)   & SU(3)         \\
\hline
\multicolumn{7}{c}{--- coupling constants ---}                                                           \\
$g_{\sigma N}$             & 9.57   & 9.57           & 8.78   & 8.78           & 10.217  & 10.217        \\
$g_{2}$ (fm$^{-1}$)        & 12.28  & 12.28          & 27.88  & 27.88          & 10.431  & 10.431        \\
$g_{3}$                    & -8.98  & -8.98          & -14.40 & -14.40         & -28.885 & -28.885       \\
$g_{\omega N}$             & 10.61  & 10.26          & 8.71   & 8.43           & 12.868  & 12.450        \\
$g_{\rho N}$               & 4.10   & 4.10           & 4.27   & 4.27           & 4.474   & 4.474         \\
$g_{\phi N}$               & ---    & -3.50          & ---    & -2.88          & ---     & -4.250        \\
$g_{\sigma\Lambda}$        & 5.84   & 7.25           & 5.32   & 6.51           & 6.269   & 7.853         \\
$g_{\sigma\Sigma}$         & 3.87   & 5.28           & 2.85   & 4.04           & 4.709   & 6.293         \\
$g_{\sigma\Xi}$            & 3.06   & 5.87           & 2.83   & 5.20           & 3.242   & 6.408         \\
$g_{\sigma^{\ast}\Lambda}$ & 3.73   & 2.60           & 2.03   & 1.95           & 5.374   & 4.174         \\
$g_{\sigma^{\ast}\Xi}$     & 9.67   & 6.82           & 7.89   & 5.55           & 11.765  & 8.378         \\
\multicolumn{7}{c}{--- properties of symmetric nuclear matter ---}                                       \\
$n_{B}^{0}$ (fm$^{-3}$)    & 0.153  & 0.153          & 0.153  & 0.153          & 0.148   & 0.148         \\
$w_{0}$ (MeV)              & -16.3  & -16.3          & -16.3  & -16.3          & -16.299 & -16.299       \\
$M_{N}^{\ast}/M_{N}$       & 0.70   & 0.70           & 0.78   & 0.78           & 0.60    & 0.60          \\
$K_{v}$ (MeV)              & 300    & 300            & 240    & 240            & 271.76  & 271.76        \\
$a_{4}$ (MeV)              & 32.5   & 32.5           & 32.5   & 32.5           & 37.4    & 37.4          \\
$L$ (MeV)                  & 93.9   & 93.9           & 89.7   & 89.7           & 118.0   & 118.0         \\
\end{tabular}
\end{ruledtabular}
\end{table*}
\begin{table*}
\caption{\label{tab:cc-non-linear-sigma-omega-rho}
Coupling constants and properties of symmetric nuclear matter in the TM1, FSUGold and IU-FSU models.  
The relations, $g_{\sigma^{\ast}N}=0$ and $g_{\sigma^{\ast} \Lambda} = g_{\sigma^{\ast} \Sigma}$, are assumed.  
For the TM1 model, we take $M_{N}=938$ MeV and $m_{\sigma}=511.198$ MeV~\cite{Sugahara:1993wz}, 
while, for the FSUGold and IU-FSU models, $m_{\sigma}=491.500$ MeV, $m_{\omega}=782.500$ MeV and $m_{\rho}=763.000$ MeV~\cite{ToddRutel:2005zz,Fattoyev:2010mx}.  
The other masses are the same as in Table~\ref{tab:cc-QMC-and-CQMC}.  
}
\begin{ruledtabular}
\begin{tabular}{lcccccc}
\                          & \multicolumn{2}{c}{TM1}   & \multicolumn{2}{c}{FSUGold}   & \multicolumn{2}{c}{IU-FSU} \\
vector sym.                & SU(6)      & SU(3)        & SU(6)       & SU(3)       & SU(6)        & SU(3)           \\
\hline
\multicolumn{7}{c}{--- coupling constants ---}                                                                      \\
$g_{\sigma N}$             & 10.029     & 10.029       & 10.592      & 10.592      & 9.971        & 9.971           \\
$g_{2}$ (fm$^{-1}$)        & 7.233      & 7.233        & 4.277       & 4.277       & 8.493        & 8.493           \\
$g_{3}$                    & 0.618      & 0.618        & 49.856      & 49.856      & 0.488        & 0.488           \\
$c_{3}$                    & 71.308     & 81.601       & 418.394     & 522.820     & 144.220      & 171.586         \\
$\Lambda_{\omega\rho}$
\footnote{
The coupling constant, $\Lambda_{\omega\rho}$, also varies, because it is defined by $\Lambda_{\omega\rho} = \Lambda_v g_{\rho N}^2 g_{\omega N}^2$, 
where $\Lambda_v$ takes the same value in both SU(3) and SU(6) cases.
}
                           & ---        & ---          & 212.427     & 168.100     & 360.714      & 248.010         \\
$g_{\omega N}$             & 12.614     & 12.199       & 14.302      & 13.874      & 13.032       & 12.615          \\
$g_{\rho N}$               & 4.632      & 4.640        & 5.884       & 5.395       & 6.795        & 5.821           \\
$g_{\phi N}$               & ---        & -4.164       & ---         & -4.736      & ---          & -4.306          \\
$g_{\sigma\Lambda}$        & 6.170      & 7.733        & 6.501       & 8.295       & 6.090        & 7.680           \\
$g_{\sigma\Sigma}$         & 4.472      & 6.035        & 4.820       & 6.615       & 4.517        & 6.107           \\
$g_{\sigma\Xi}$            & 3.202      & 6.328        & 3.366       & 6.953       & 3.154        & 6.334           \\
$g_{\sigma^{\ast}\Lambda}$ & 5.015      & 3.691        & 5.994       & 4.458       & 5.476        & 4.204           \\
$g_{\sigma^{\ast}\Xi}$     & 11.516     & 8.100        & 13.071      & 9.147       & 11.915       & 8.437           \\
\multicolumn{7}{c}{--- properties of symmetric nuclear matter ---}                                                  \\
$n_{B}^{0}$ (fm$^{-3}$)    & 0.145      & 0.145        & 0.148       & 0.148       & 0.155       & 0.155            \\
$w_{0}$ (MeV)              & -16.3      & -16.3        & -16.30      & -16.30      & -16.40      & -16.40           \\
$M_{B}^{\ast}/M_{B}$       & 0.634      & 0.634        & 0.61        & 0.61        & 0.61        & 0.61             \\
$K_{v}$ (MeV)              & 281        & 284          & 230         & 252         & 231.2       & 237.7            \\
$a_{4}$ (MeV)              & 36.9       & 36.9         & 32.59       & 32.59       & 31.30       & 31.30            \\
$L$ (MeV)                  & 110.9      & 110.8        & 60.3        & 66.6        & 47.2        & 54.6             \\
\end{tabular}
\end{ruledtabular}
\end{table*}

\section{Numerical results and discussions}
\label{sec:results-and-discussions}
\subsection{Properties of symmetric nuclear matter}
\label{subsec:matter-properties}

As seen in Tables~\ref{tab:cc-QMC-and-CQMC}, \ref{tab:cc-non-linear-sigma} and \ref{tab:cc-non-linear-sigma-omega-rho}, 
the properties of symmetric nuclear matter are well reproduced in all the models.  

For the QHD-type models, the results calculated by GM1, GM3 and NL3 are presented in Table~\ref{tab:cc-non-linear-sigma}, 
and those by TM1, FSUGold and IU-FSU are in Table~\ref{tab:cc-non-linear-sigma-omega-rho}.  
In the former group, the NL potential involves the self-interaction terms of the $\sigma$ meson, 
while, in the latter group, in addition, the NL terms of the vector ($\omega$ and $\vec{\rho\,}$) mesons are taken into account.  
In the GM1, GM3 and NL3 models, the same saturation properties can be achieved in both SU(6) and SU(3) cases, 
as explained in section~\ref{subsec:SU(3)} (see also Eq.~(\ref{eq:EOM-omega}) and (\ref{eq:EOM-phi})).  
In contrast, in the TM1, FSUGold and IU-FSU models, 
the symmetry energy, incompressibility and slope parameter, $L$, in SU(3) symmetry are slightly changed from the original values (given in SU(6) symmetry), 
because the terms of $c_{3}$ and/or $\Lambda_{\omega \rho}$ in the NL potential, 
which has the quartic dependence of the nuclear density in the energy density and pressure of matter, also take part in reproducing the saturation condition.  

Furthermore, we notice the following two points.  
Firstly, in the extension of SU(6) to SU(3) symmetry, the coupling, $g_{\omega N}$, becomes smaller, 
because the (total) repulsive force is attributed not only to the $\omega$ but also to the $\phi$, which is caused by the mixing in Eq.(\ref{eq:mixing}).  
We note that the coupling constant, $g_{\phi N}$, is negative, 
because the mean-field value of the $\phi$ meson has a negative sign (see Figs.~\ref{fig:Field-QMC} - \ref{fig:Field-FSU}).  

Secondly, the coupling, $g_{\sigma^\ast Y}$, (or the $\sigma^{\ast}$ field itself) in SU(3) symmetry is suppressed in all the models, comparing with that in the SU(6) case.  
In contrast, the $\sigma$-$Y$ couplings in SU(3) symmetry are more enhanced than in SU(6) symmetry.  
This enhancement may counterbalance the additional, repulsive force due to the $\phi$ meson in the $Y$-$N$ interaction, 
because the (total) repulsive force in the SU(3) case is stronger than in the SU(6) case.  

\subsection{Neutron Stars}
\label{subsec:neutron-star}

%
\begin{figure}
\includegraphics[width=200pt,keepaspectratio,clip,angle=270]{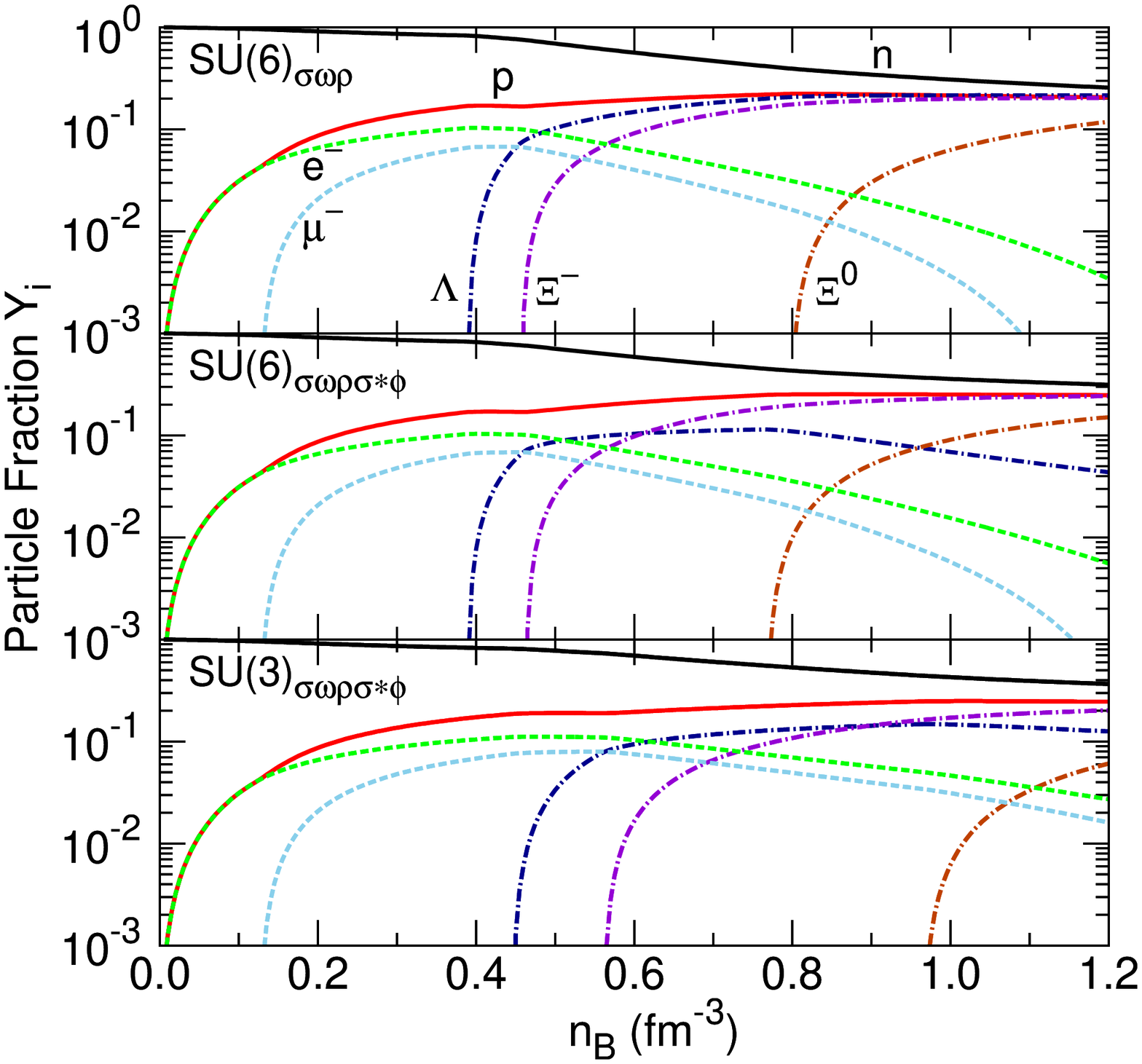}%
\includegraphics[width=200pt,keepaspectratio,clip,angle=270]{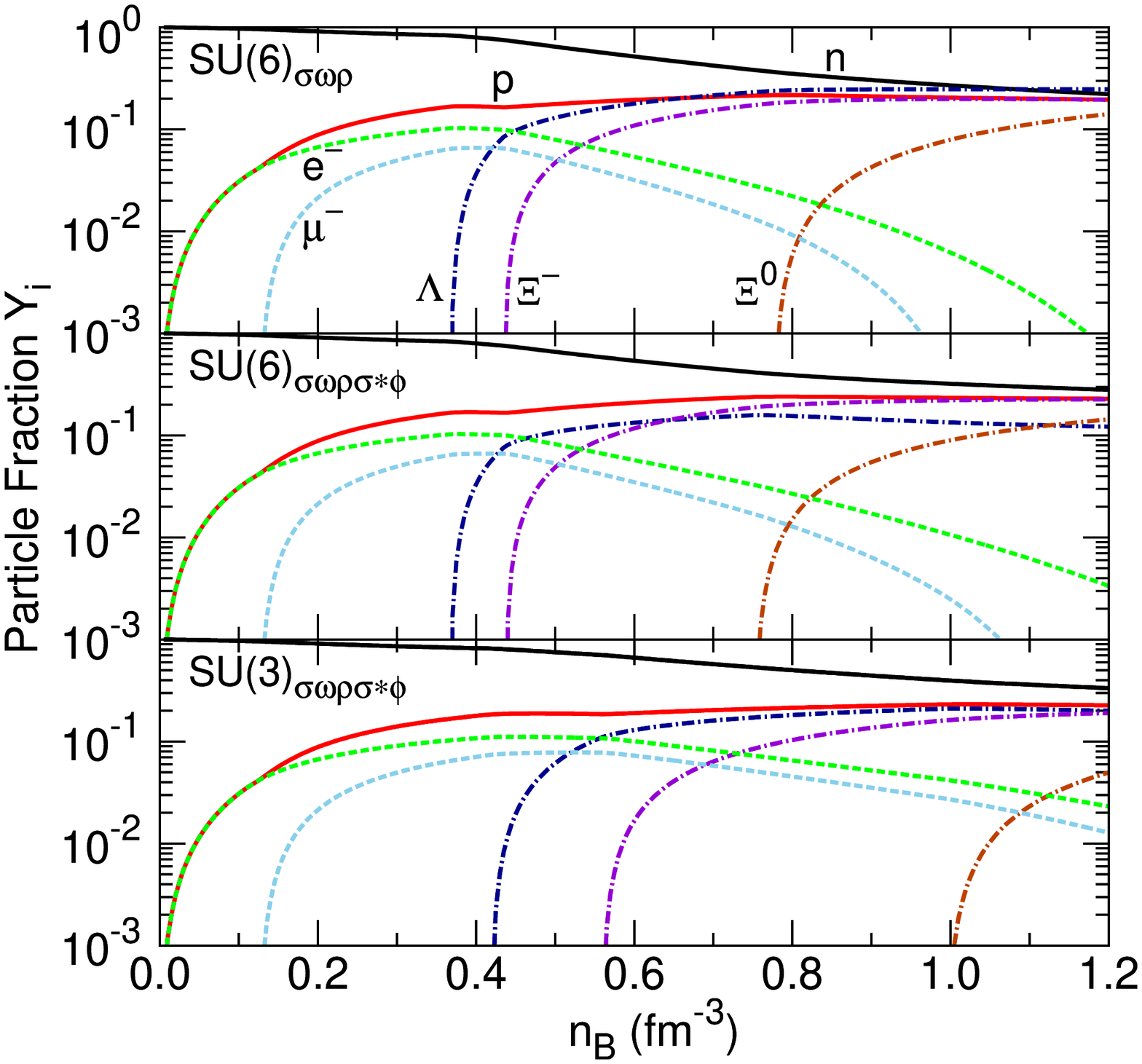}%
\caption{\label{fig:Composition-QMC} Particle fractions, $Y_{i}$, in the QMC and CQMC models (left: QMC, right: CQMC). }
\end{figure}
\begin{figure}
\includegraphics[width=200pt,keepaspectratio,clip,angle=270]{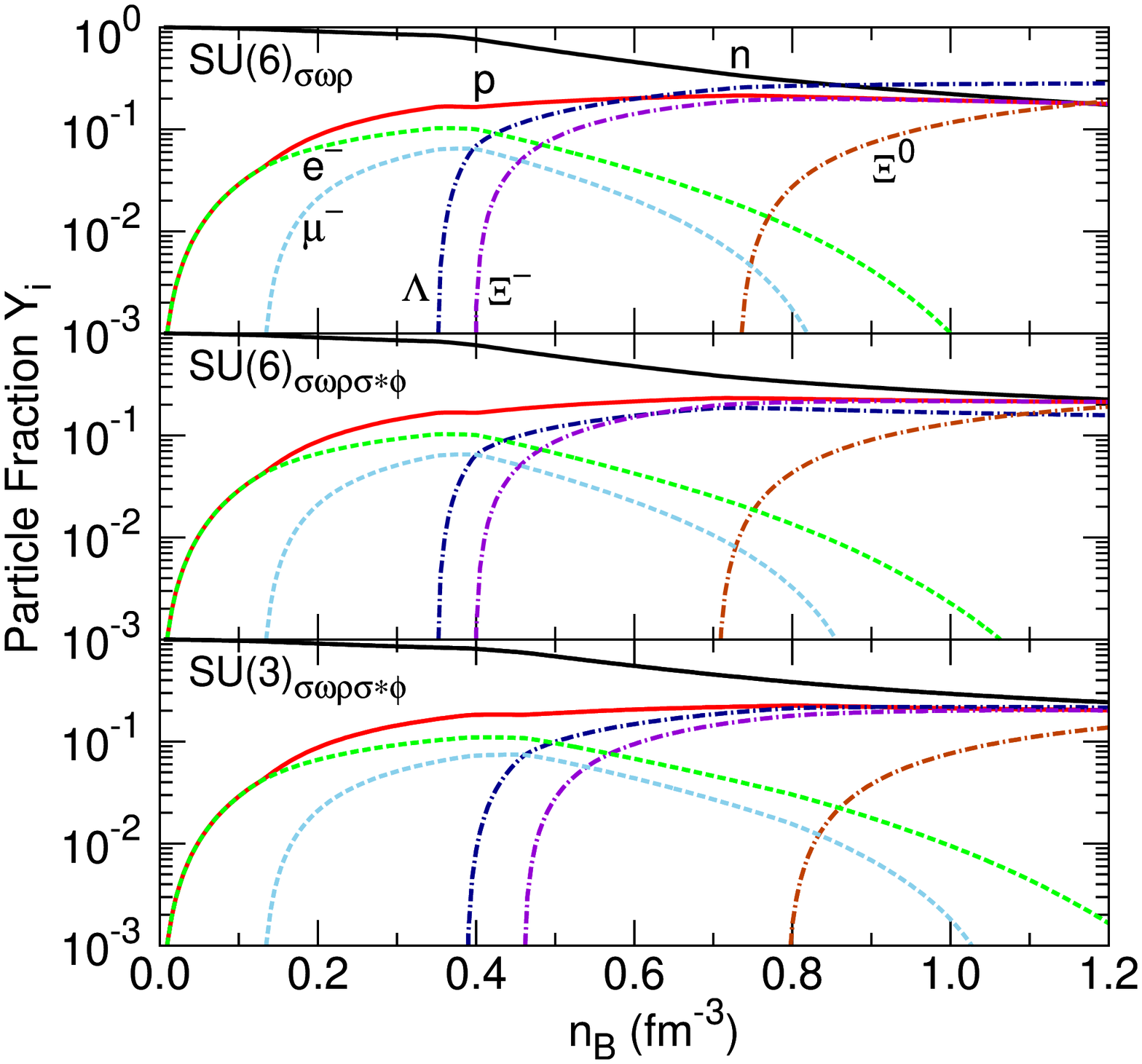}%
\includegraphics[width=200pt,keepaspectratio,clip,angle=270]{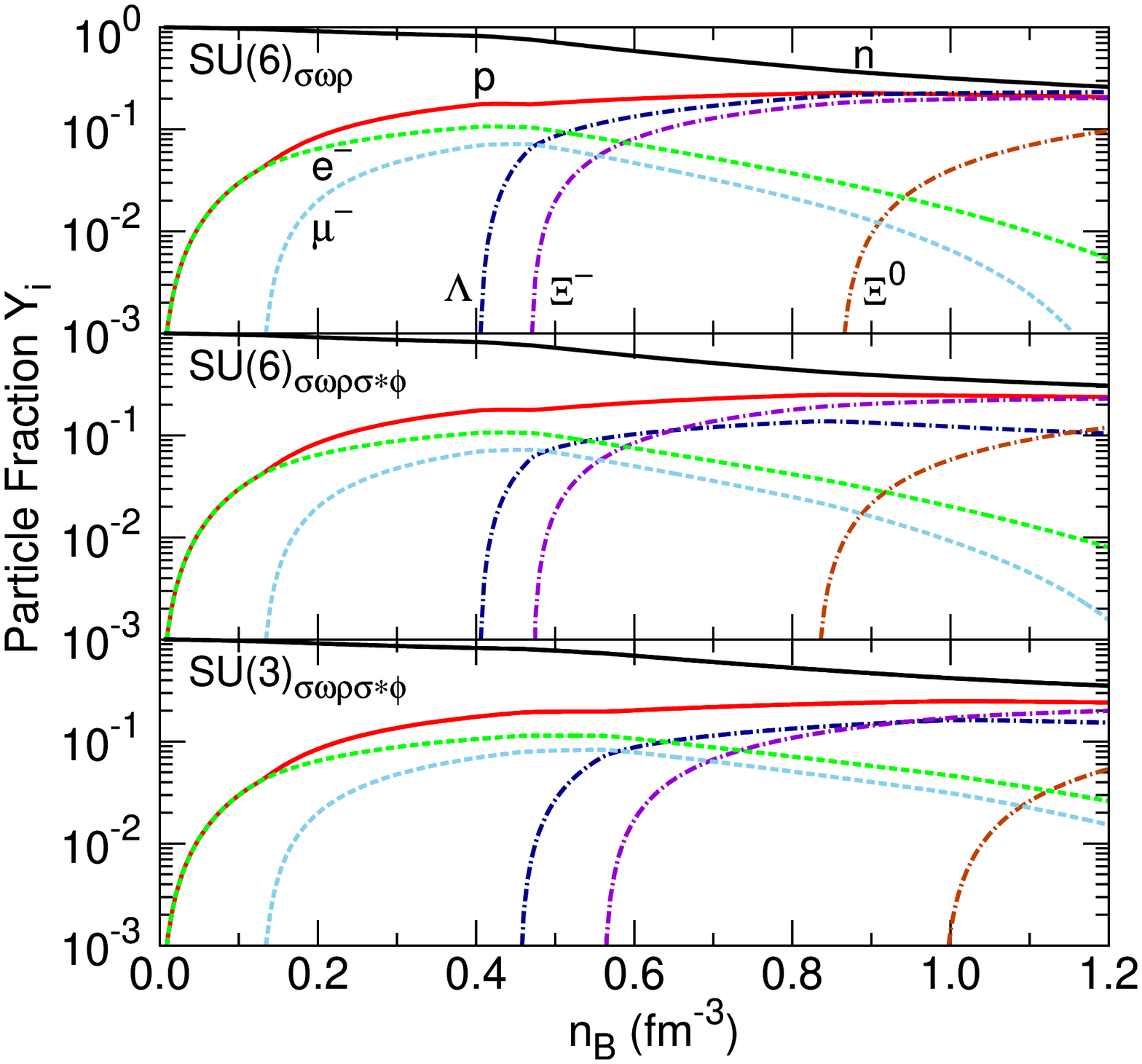} \\ %
\includegraphics[width=200pt,keepaspectratio,clip,angle=270]{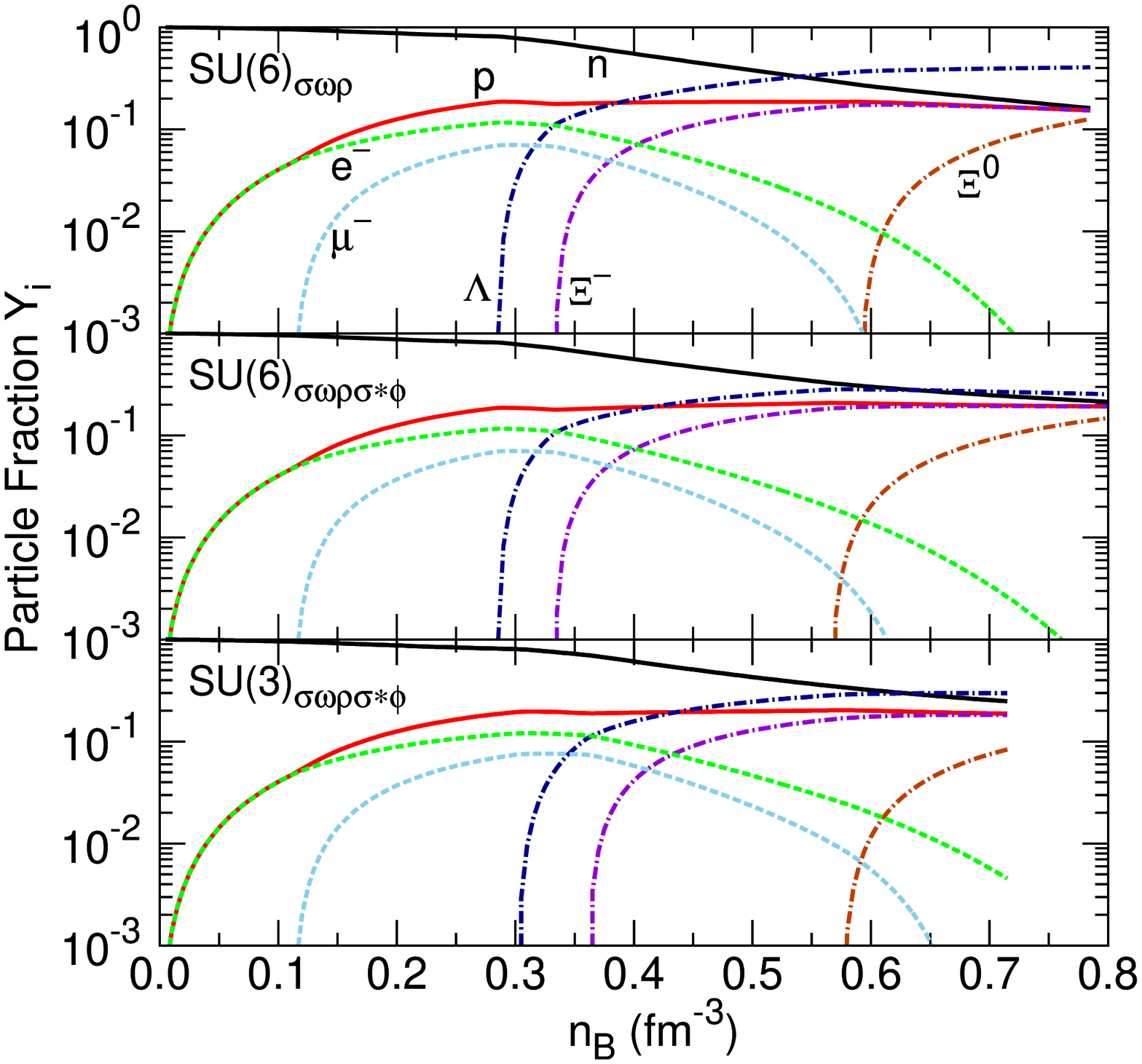}%
\includegraphics[width=200pt,keepaspectratio,clip,angle=270]{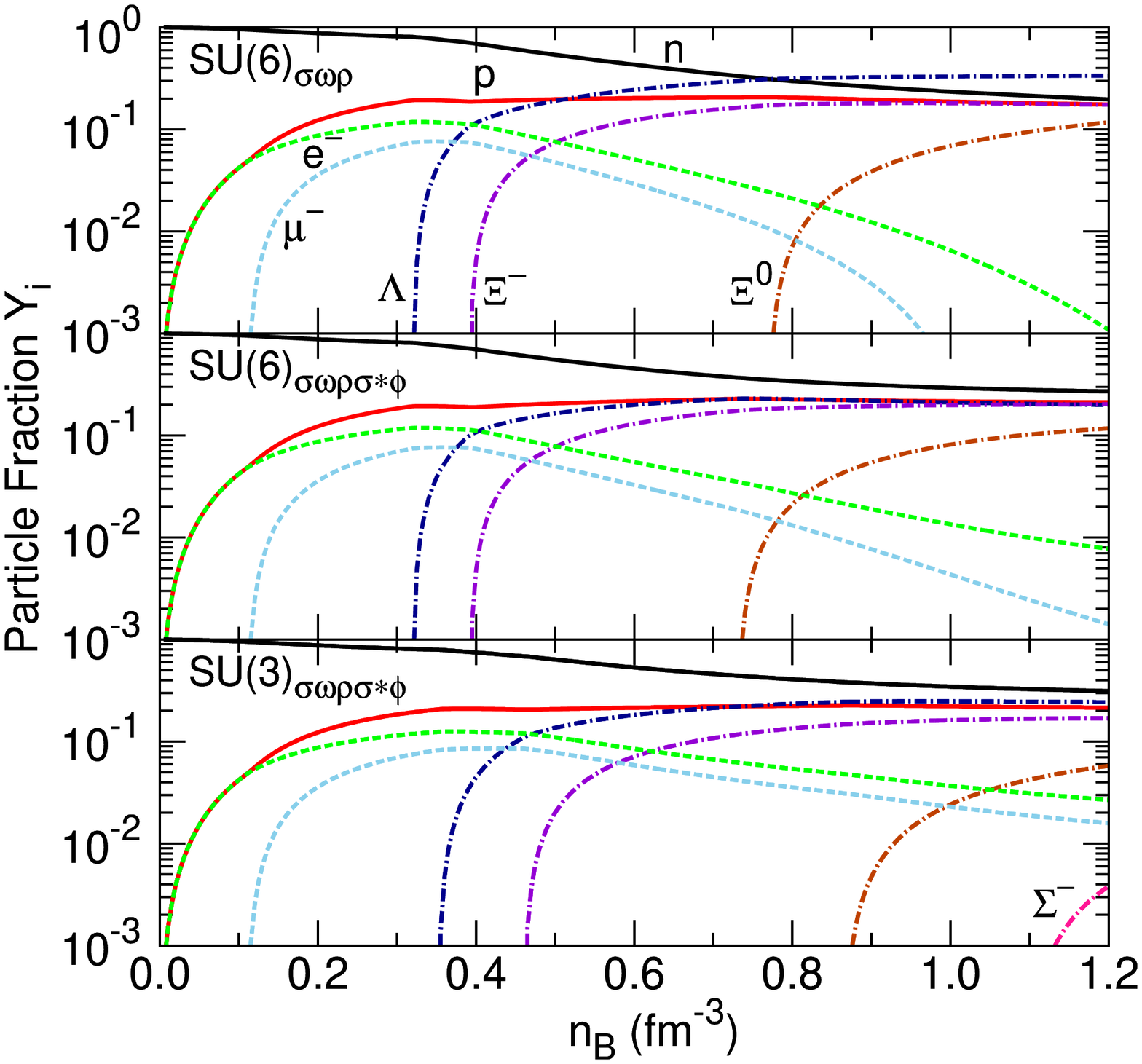}%
\caption{\label{fig:Composition-GM1} Particle fractions, $Y_{i}$, in the GM1, GM3, NL3 and TM1 models (upper left: GM1, upper right: GM3, lower left: NL3, lower right: TM1). }
\end{figure}
\begin{figure}
\includegraphics[width=200pt,keepaspectratio,clip,angle=270]{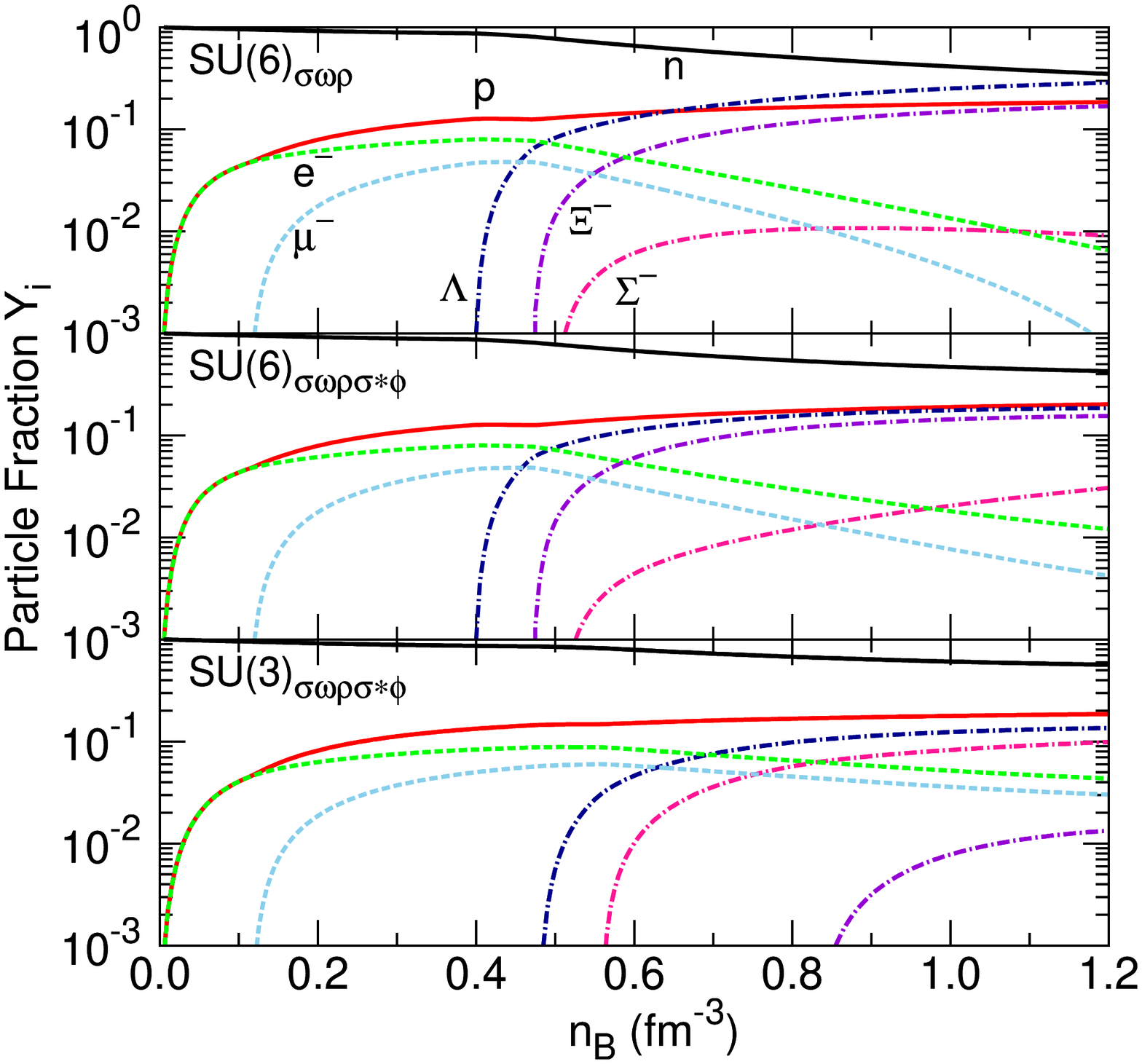}%
\includegraphics[width=200pt,keepaspectratio,clip,angle=270]{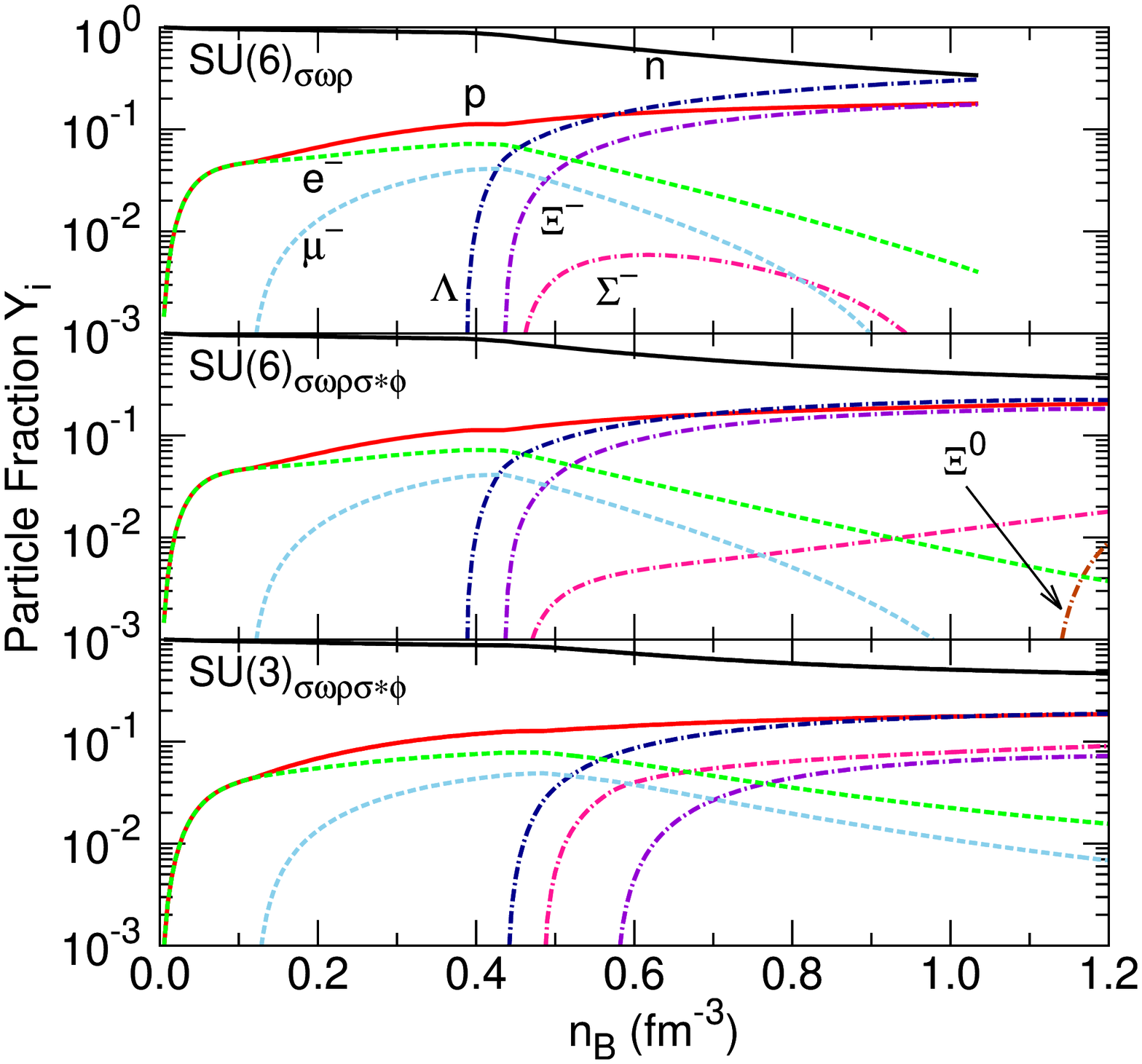}%
\caption{\label{fig:Composition-FSU} Particle fractions, $Y_{i}$, in the FSUGold and IU-FSU models (left: FSUGold, right: IU-FSU).}
\end{figure}
%
%
\begin{figure}
\includegraphics[width=200pt,keepaspectratio,clip,angle=270]{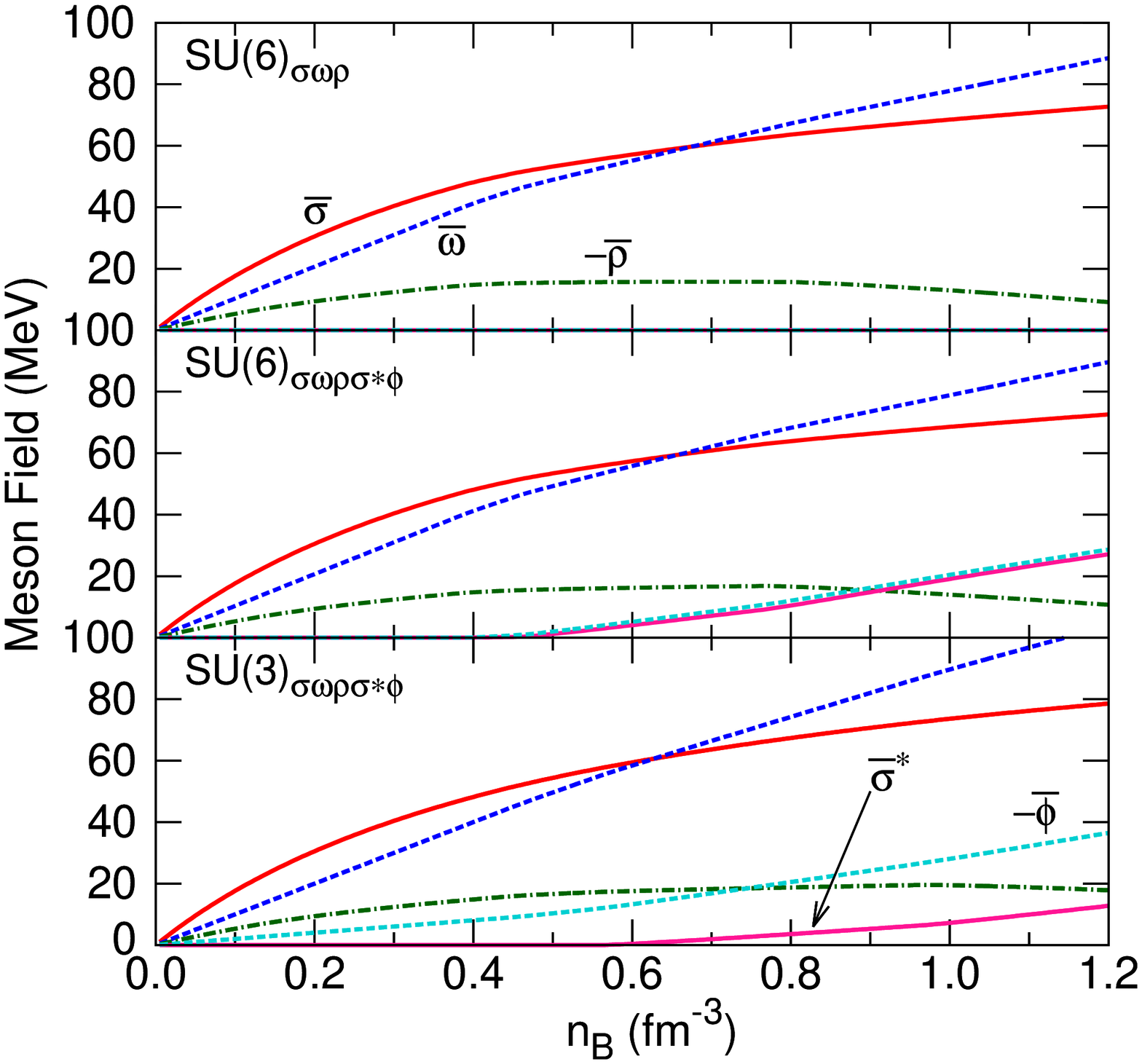}%
\includegraphics[width=200pt,keepaspectratio,clip,angle=270]{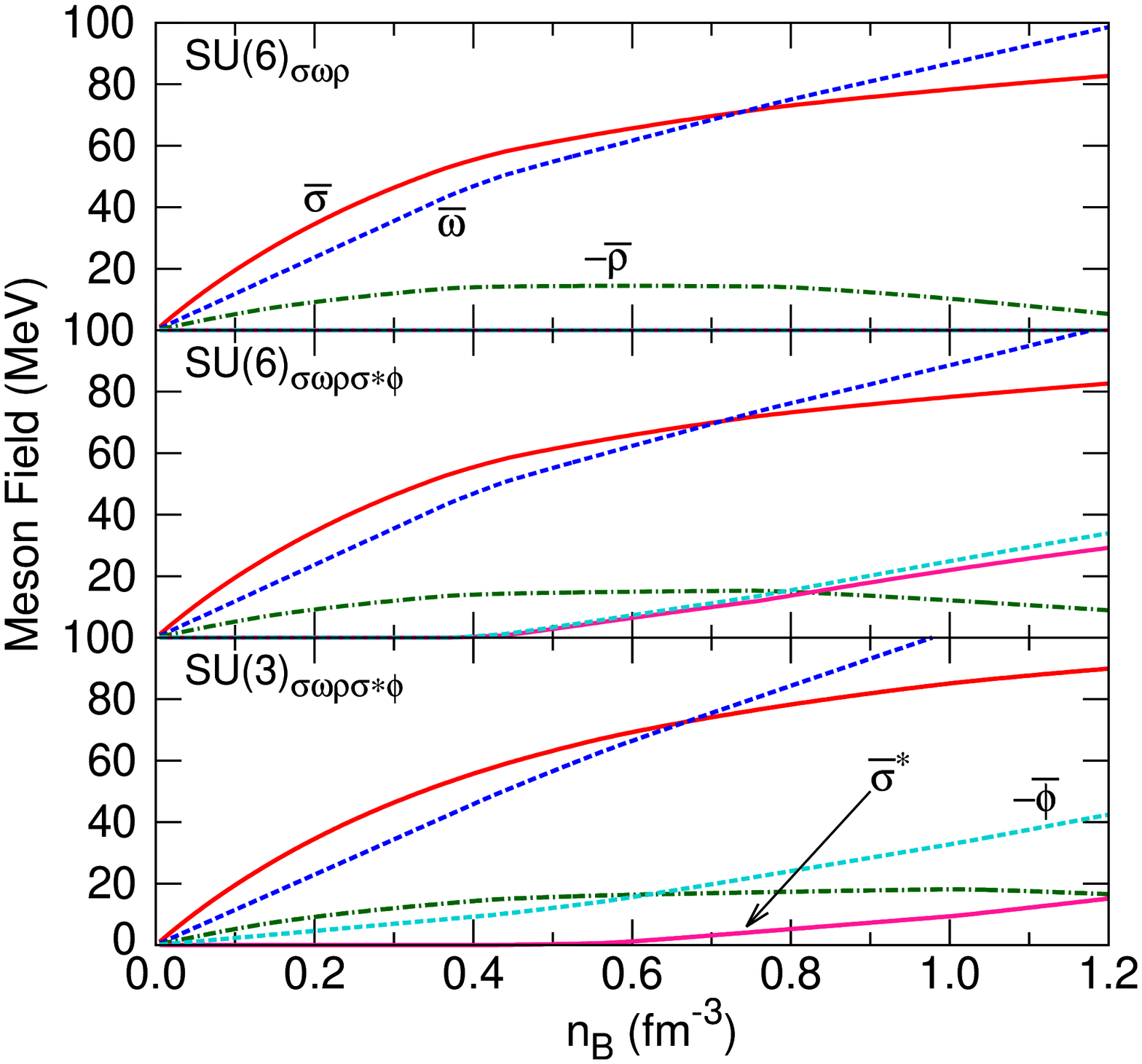}%
\caption{\label{fig:Field-QMC} Meson fields in the QMC and CQMC models (left: QMC, right: CQMC).}
\end{figure}
\begin{figure}
\includegraphics[width=200pt,keepaspectratio,clip,angle=270]{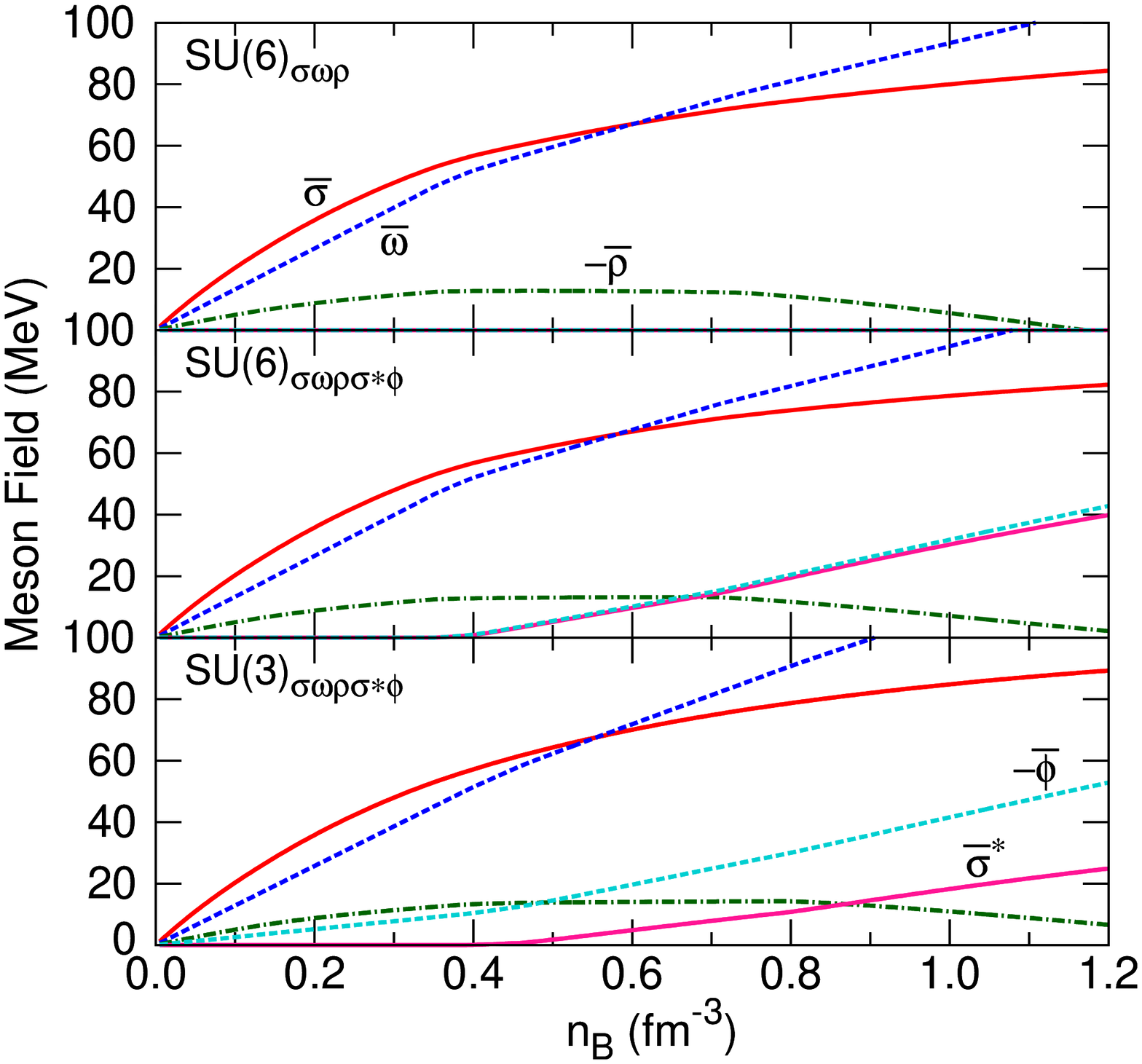}%
\includegraphics[width=200pt,keepaspectratio,clip,angle=270]{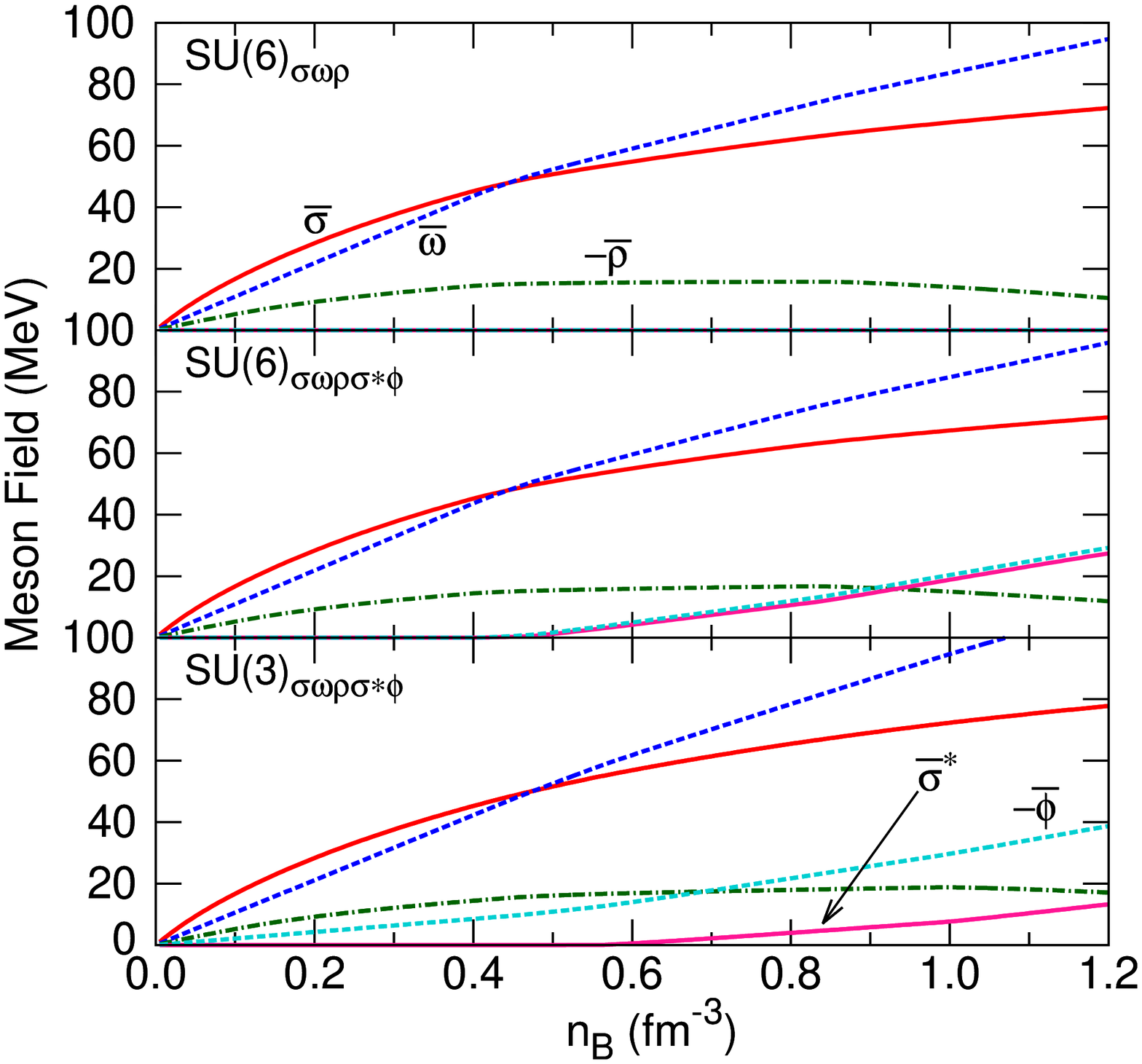} \\ %
\includegraphics[width=200pt,keepaspectratio,clip,angle=270]{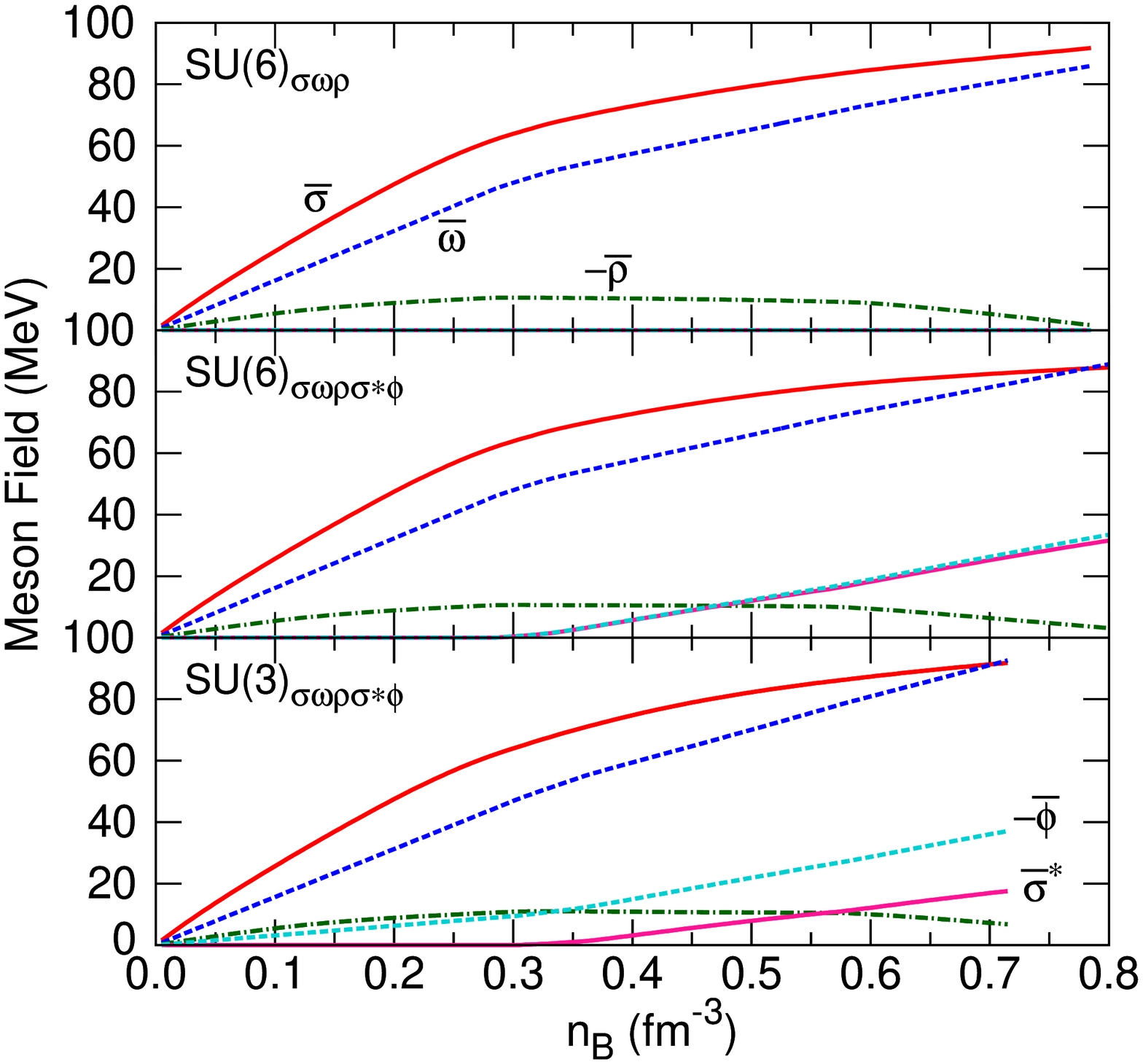}%
\includegraphics[width=200pt,keepaspectratio,clip,angle=270]{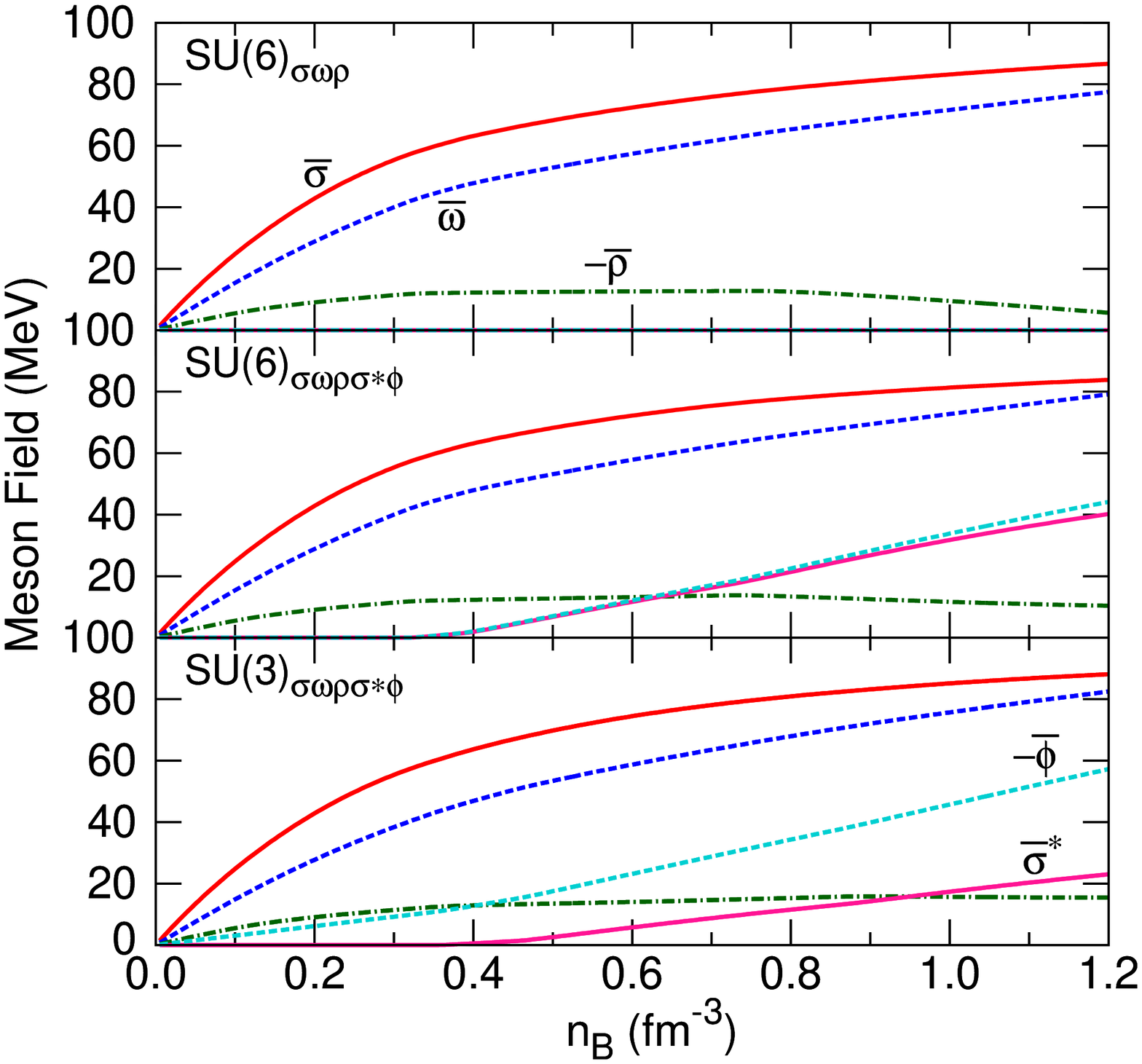}%
\caption{\label{fig:Field-GM1} Meson fields in the GM1, GM3, NL3 and TM1 models (upper left: GM1, upper right: GM3, lower left: NL3, lower right: TM1).}
\end{figure}
\begin{figure}
\includegraphics[width=200pt,keepaspectratio,clip,angle=270]{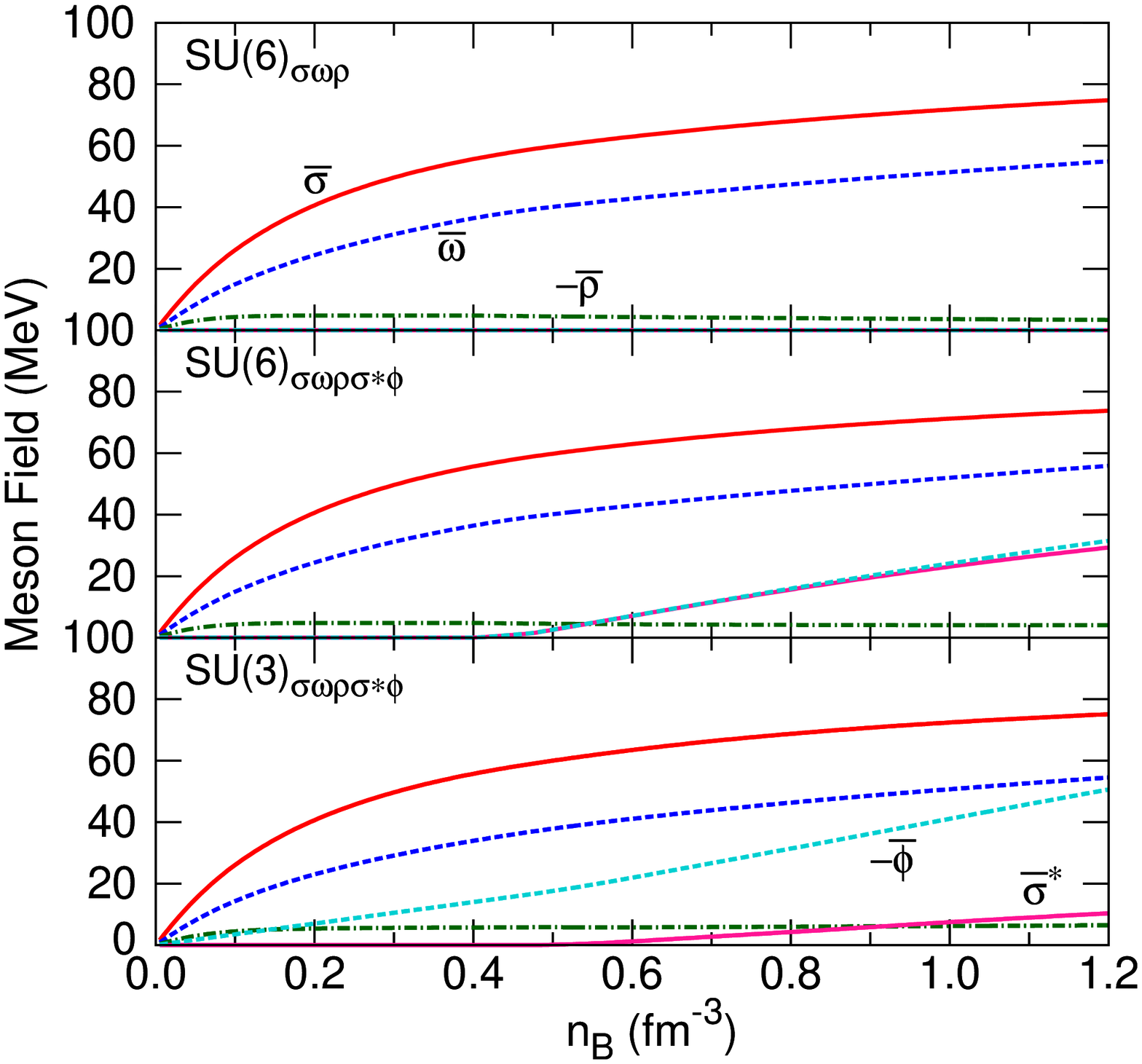}%
\includegraphics[width=200pt,keepaspectratio,clip,angle=270]{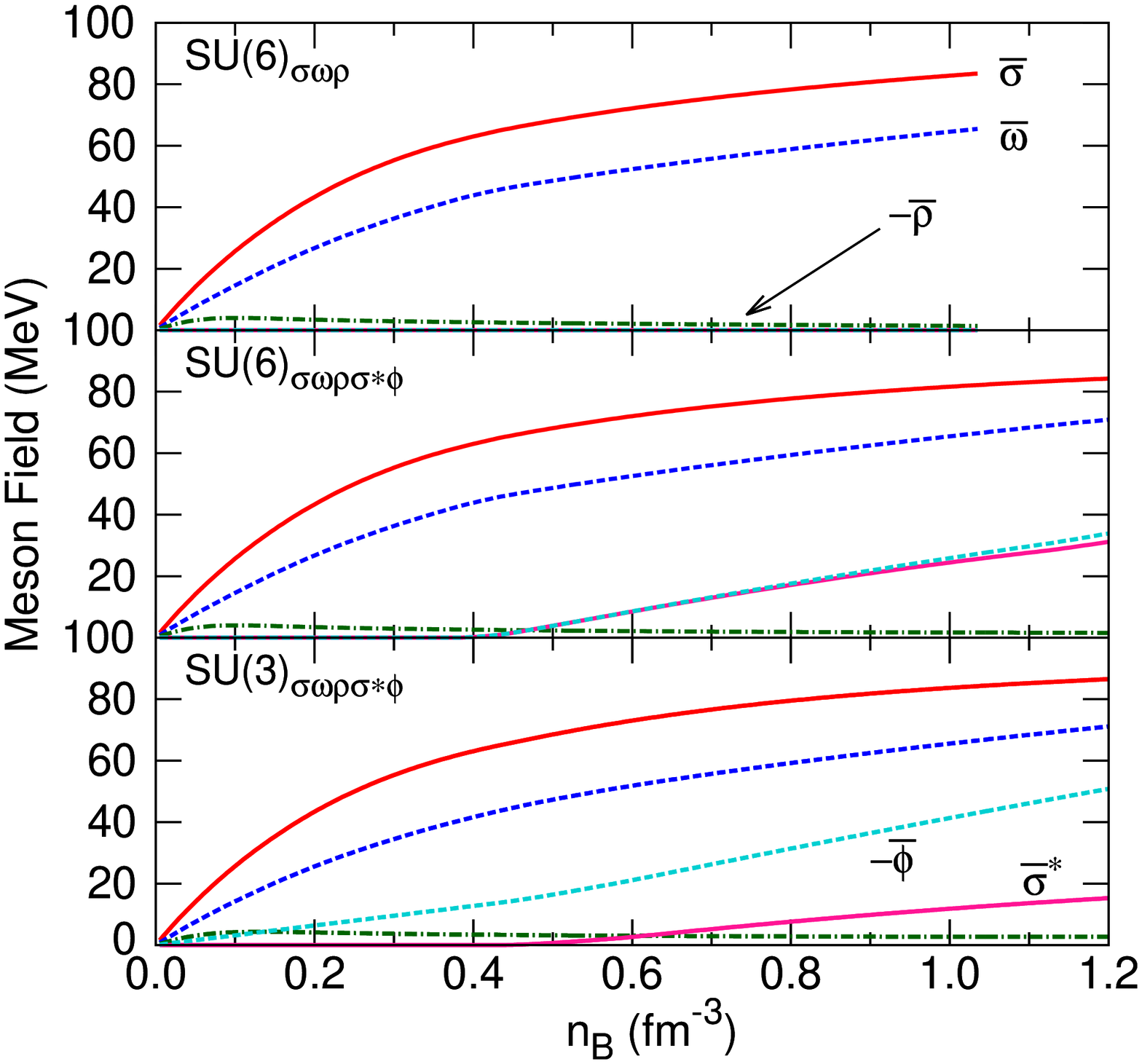}%
\caption{\label{fig:Field-FSU} Meson fields in the FSUGold and IU-FSU models (left: FSUGold, right: IU-FSU).}
\end{figure}
%
%
\begin{figure}
\includegraphics[width=200pt,keepaspectratio,clip,angle=270]{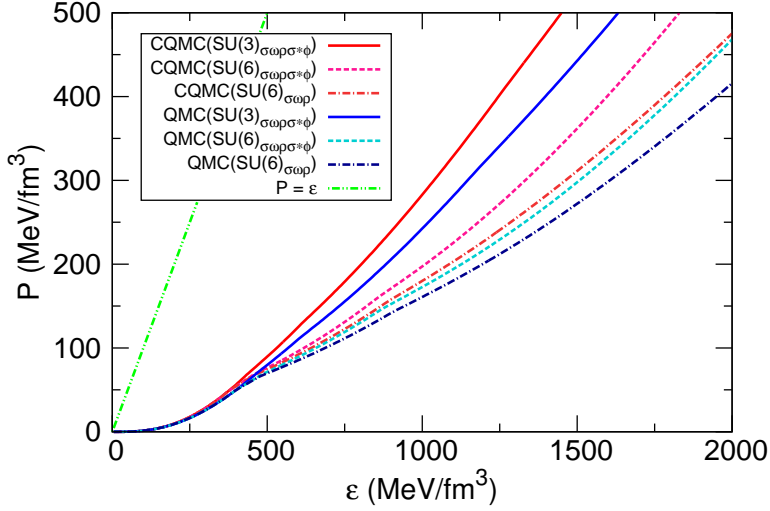}%
\caption{\label{fig:EOS-QMC-CQMC} Equations of state in the QMC and CQMC models.}
\end{figure}
\begin{figure}
\includegraphics[width=160pt,keepaspectratio,clip,angle=270]{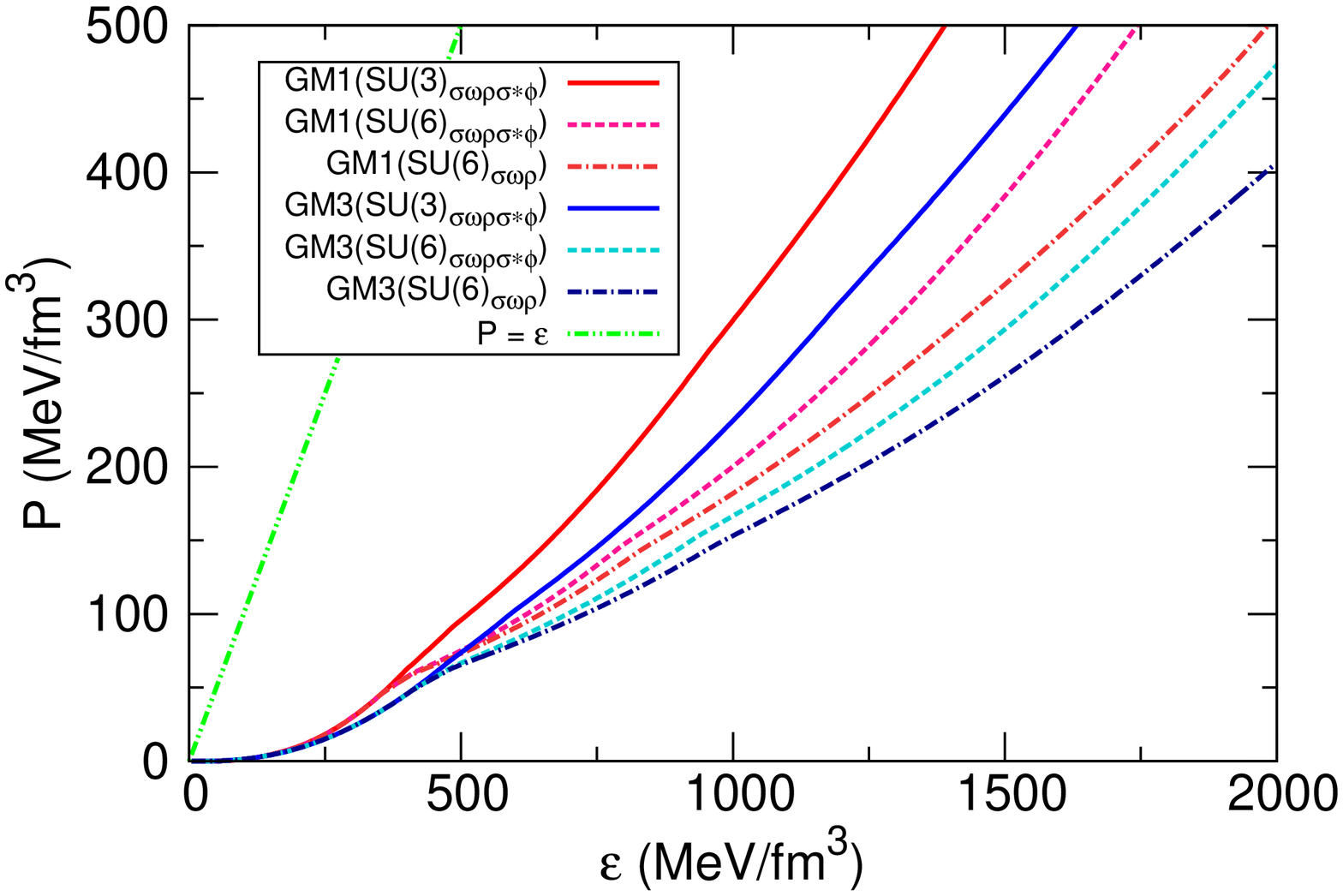}%
\includegraphics[width=160pt,keepaspectratio,clip,angle=270]{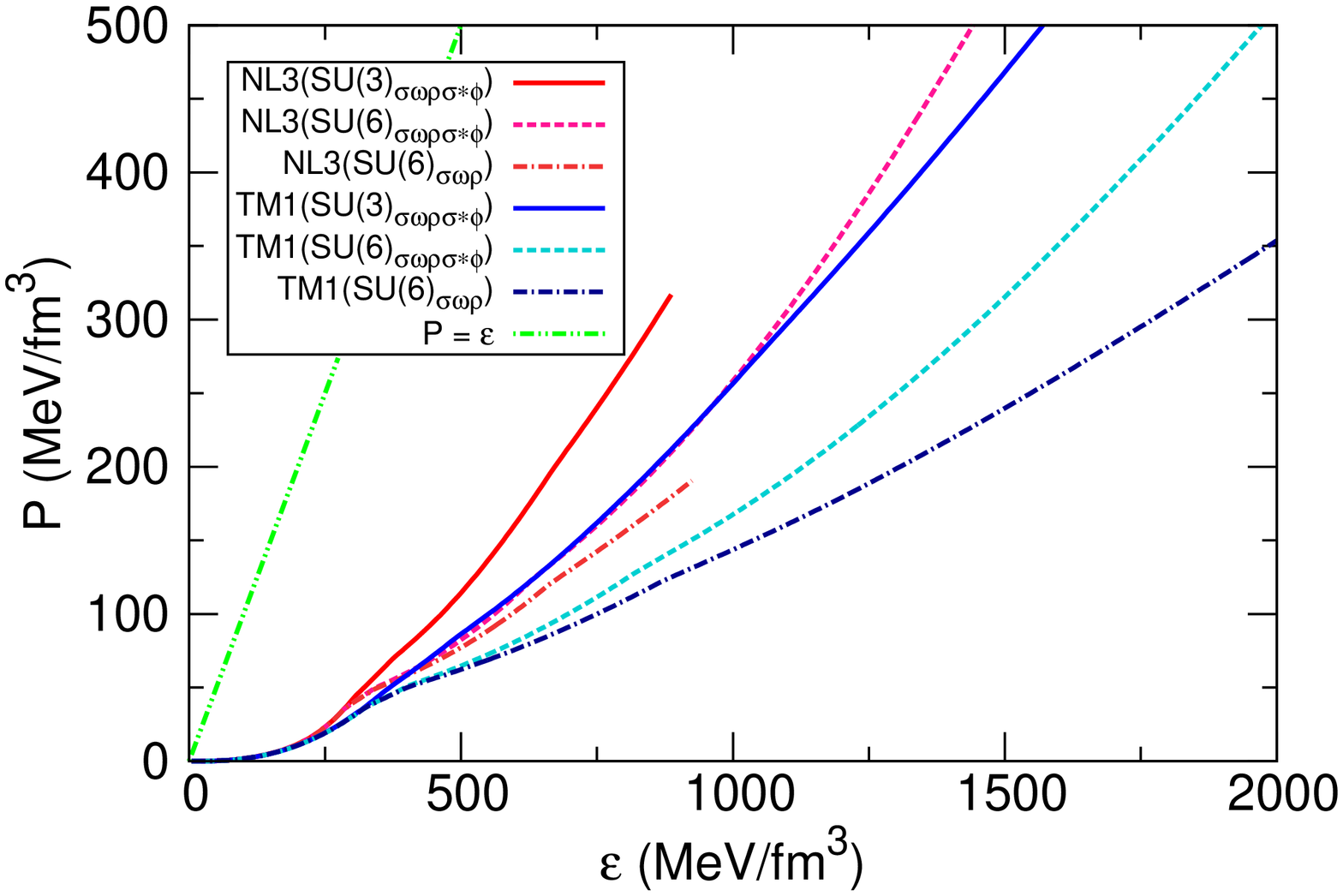}%
\caption{\label{fig:EOS-GM1-GM3} Equations of state in the GM1, GM3, NL3 and TM1 models (left: GM1 and GM3, right: NL3 and TM1).}
\end{figure}
\begin{figure}
\includegraphics[width=200pt,keepaspectratio,clip,angle=270]{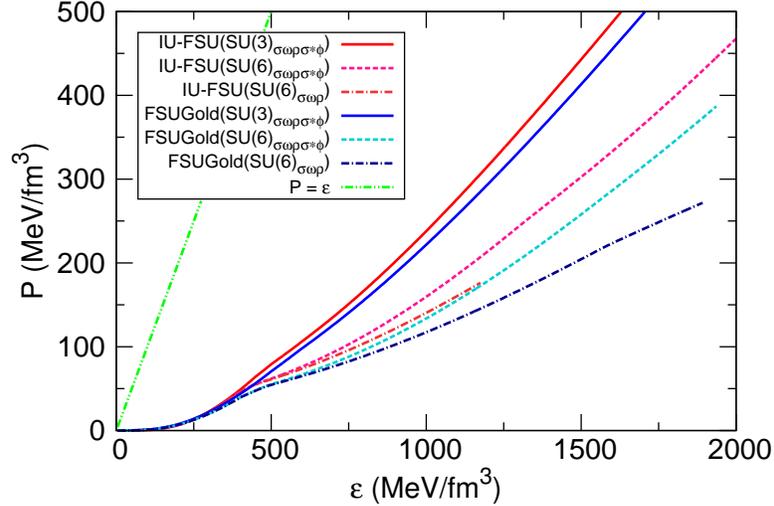}%
\caption{\label{fig:EOS-FSU-IU-FSU} Equations of state in the FSUGold and IU-FSU models.}
\end{figure}
%
%
\begin{figure}
\includegraphics[width=200pt,keepaspectratio,clip,angle=270]{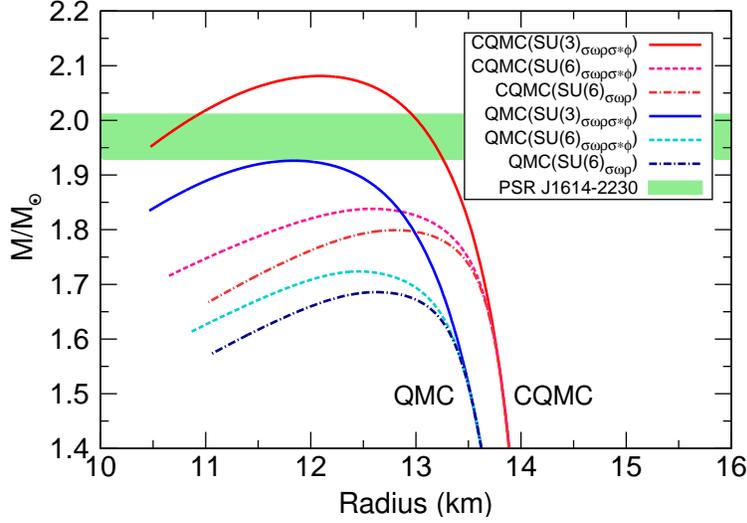}%
\caption{\label{fig:TOV-QMC-CQMC} Mass-radius relations in the QMC and CQMC models.}
\end{figure}
\begin{figure}
\includegraphics[width=160pt,keepaspectratio,clip,angle=270]{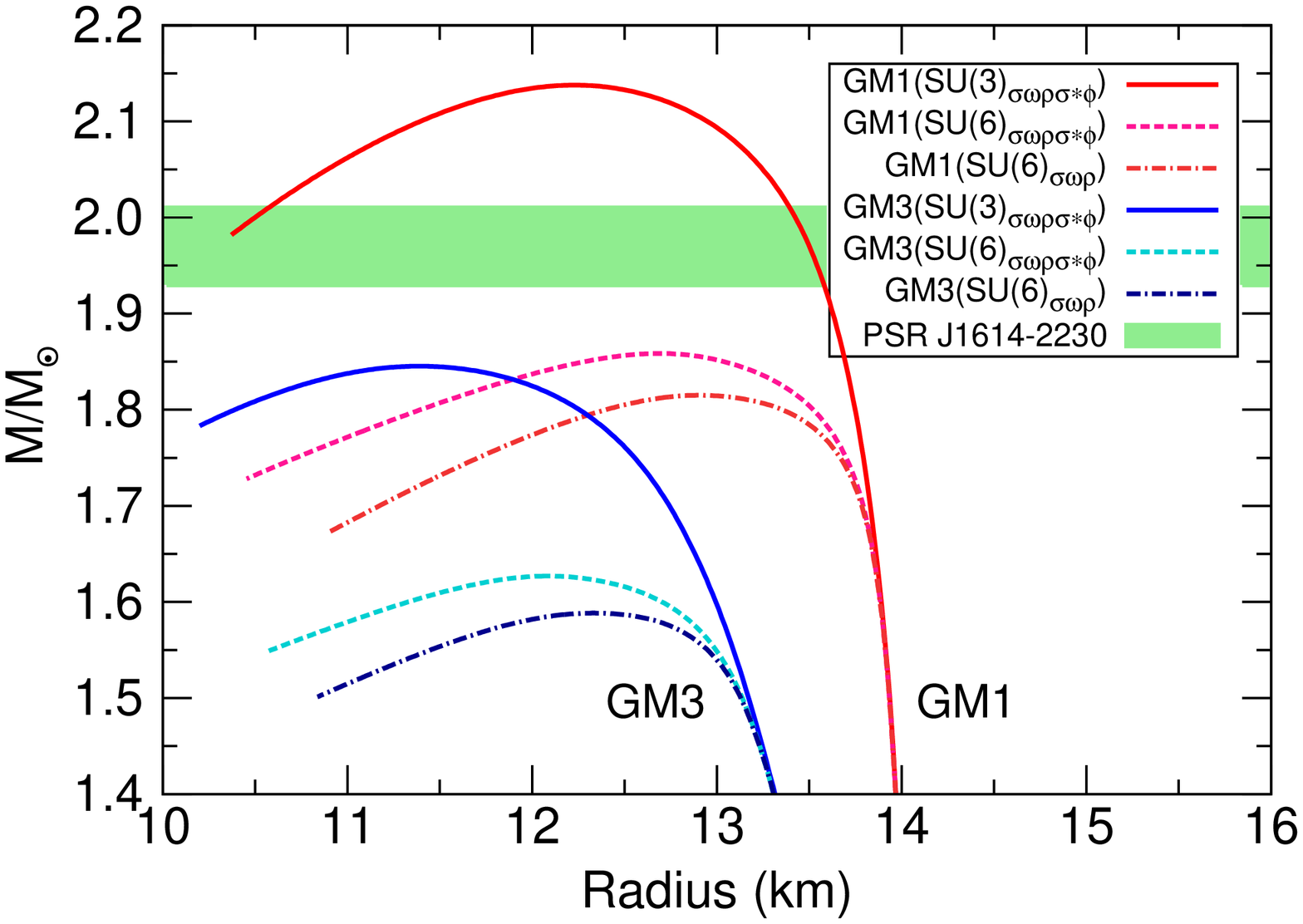}%
\includegraphics[width=160pt,keepaspectratio,clip,angle=270]{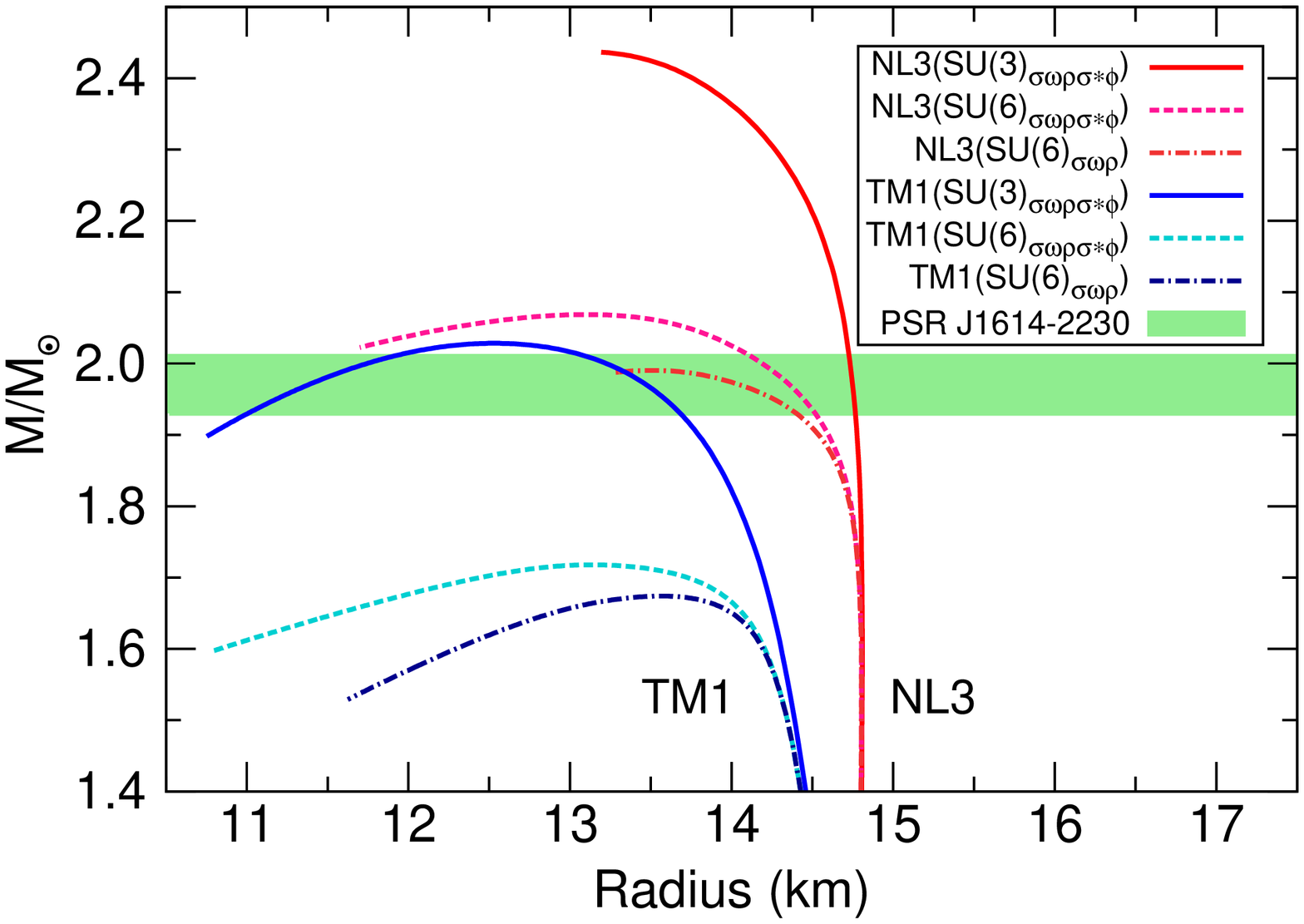}%
\caption{\label{fig:TOV-GM1-GM3} Mass-radius relations in the GM1, GM3, NL3 and TM1 models (left: GM1 and GM3, right: NL3 and TM1). 
In the NL3 model, the (red) solid curve for SU(3) symmetry does not yet reach the maximum point because the nucleon mass in matter becomes negative at the endpoint. }
\end{figure}
\begin{figure}
\includegraphics[width=200pt,keepaspectratio,clip,angle=270]{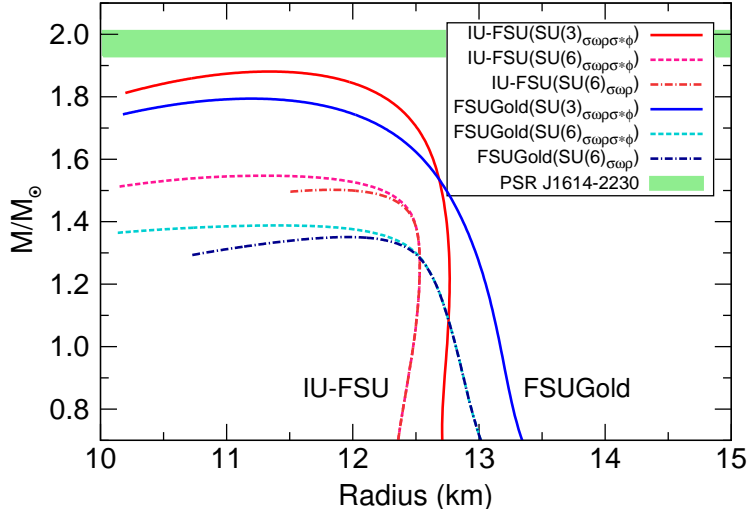}%
\caption{\label{fig:TOV-FSU-IU-FSU} Mass-radius relations in the FSUGold and IU-FSU models.}
\end{figure}

\begin{table}
\caption{\label{tab:NS-properties}
Properties of a neutron star in SU(6) or SU(3) symmetry. 
We list the neutron-star radius, $R_{\max}$ (in km), the ratio of the neutron-star mass to the solar mass, $M_{\max}/M_{\odot}$, 
and the central density, $n_{c}$ (in fm$^{-3}$) at the maximum-mass point. 
In these calculations,  we consider all the mesons ($\sigma$, $\omega$, $\rho$, $\sigma^{\ast}$ and $\phi$).  
}
\begin{ruledtabular}
\begin{tabular}{lcccccc}
\       & \multicolumn{3}{c}{SU(6)}                   & \multicolumn{3}{c}{SU(3)}                          \\
\       & $R_{\max}$ & $M_{\max}/M_{\odot}$ & $n_{c}$ & $R_{\max}$ & $M_{\max}/M_{\odot}$ & $n_{c}$        \\
\hline
QMC     & 12.5       & 1.72                 & 0.85    & 11.8       & 1.93                 & 0.96           \\
CQMC    & 12.6       & 1.84                 & 0.84    & 12.1       & 2.08                 & 0.90           \\
GM1     & 12.7       & 1.86                 & 0.82    & 12.2       & 2.14                 & 0.87           \\
GM3     & 12.1       & 1.63                 & 0.93    & 11.4       & 1.85                 & 1.05           \\
NL3\footnote{In the NL3 model with SU(3) symmetry, the nucleon mass becomes negative before the neutron-star mass reaches the maximum point. 
Therefore, the maximum mass is not given in SU(3) symmetry. 
}    
        & 13.1       & 2.07                 & 0.78    & ---        & ---                  & ---            \\
TM1     & 13.1       & 1.72                 & 0.77    & 12.5       & 2.03                 & 0.86           \\
FSUGold & 11.4       & 1.39                 & 1.03    & 11.2       & 1.79                 & 1.08           \\
IU-FSU  & 11.3       & 1.55                 & 1.03    & 11.3       & 1.88                 & 1.02           \\
\end{tabular}
\end{ruledtabular}
\end{table}

In the core of a neutron star, 
the charge neutrality and $\beta$ equilibrium under weak processes are imposed in solving the TOV equation~\cite{Tolman:1934za,Oppenheimer:1939ne}.  
To obtain the realistic relation between the mass and radius of a neutron star, for the EoS at very low nuclear densities ($\leq 0.068$ fm$^{-3}$), 
we use the models given by Baym, Bethe, Pethick and Sutherland~\cite{Baym:1971ax,Baym:1971pw}.  
In fact, the radius is relatively sensitive to the EoS at low density.  

In Figs.~\ref{fig:Composition-QMC} - \ref{fig:Composition-FSU}, we show the particle fractions in the core of a neutron star.  
In each model, we calculate three cases: 
(1) only the non-strange mesons ($\sigma$, $\omega$ and $\rho$) are included in SU(6) symmetry; 
(2) all the mesons including the $\sigma^{\ast}$ and $\phi$ are considered in SU(6) symmetry; 
(3) all the mesons are included in SU(3) symmetry.  
We note that, in some panels in the figures, the calculation stops at a certain density because the effective nucleon mass becomes zero beyond that density.  

As seen in the figures, from case (1) to (3) in order, the hyperons are created at higher densities.  
For example, the threshold densities of the $\Lambda$ and $\Xi^{-}$ productions in SU(3) symmetry are higher than those in SU(6) symmetry, 
which makes the fractions of hyperons small at high densities and thus increases the neutron fraction.  

In the models except for FSUGold and IU-FSU, the $\Lambda$ and $\Xi^{0,-}$ hyperons are created, but the $\Sigma$ does not appear, 
because the $\Sigma$-hyperon potential in nuclear matter, $U_\Sigma^{(N)}$, is chosen to be repulsive (see section~\ref{subsec:SU(6)})\footnote{
We note that, if the Fock term is included~\cite{Miyatsu:2011bc,Katayama:2012ge,Whittenbury:2012rn}, the hyperons except the $\Xi^{-}$ disappear. }.  
However, in the FSUGold and IU-FSU models, because the rather strong, $\omega$-$\rho$ (nonlinear) repulsive interaction, 
$\Lambda_{\omega \rho} {\bar \omega}^2 {\bar \rho}^{2}$, is included (see Eq.(\ref{eq:Lagrangian-NL})), 
the $\Xi^{0,-}$ fields are very suppressed at high densities (especially, in the SU(3) case), 
and the $\Sigma^{-}$ alternatively appears beyond $n_B \simeq 0.4 - 0.6$ fm$^{-3}$.  
Furthermore, the order of the threshold densities for the $\Xi^{-}$ and $\Sigma^{-}$ is reversed in the SU(6) and SU(3) cases.  
This is a very remarkable phenomenon, 
and the isoscalar-isovector nonlinear interaction, $\Lambda_{\omega \rho} {\bar \omega}^2 {\bar \rho}^{2}$, plays a unique role in the particle fractions.  

The meson fields are presented in Figs.~\ref{fig:Field-QMC} - \ref{fig:Field-FSU}.  
As it should be, in SU(6) symmetry, the strange-meson fields appear in the density region where the hyperons are generated.  
On the other hand, in SU(3) symmetry, the $\phi$ meson contributes to the baryon interactions even at low densities because of the mixing effect.  
However, the $\sigma^{\ast}$ meson emerges above the density at which the first hyperon (usually the $\Lambda$) is created.  
This is because we assume that $g_{\sigma^{\ast} N}=0$.  
In the FSUGold and IU-FSU models, 
the fields of $\omega$ and $\rho$ mesons (especially the $\rho$) are very suppressed because of the isoscalar-isovector, nonlinear interaction.  

In Figs.~\ref{fig:EOS-QMC-CQMC} - \ref{fig:EOS-FSU-IU-FSU}, we show the EoS in each model. 
Furthermore, in Figs.~\ref{fig:TOV-QMC-CQMC} - \ref{fig:TOV-FSU-IU-FSU}, we present the mass-radius relation of a neutron star calculated by the TOV equation.  
The detail of the neutron-star properties is also shown in Table~\ref{tab:NS-properties}.  

As expected, because the isoscalar, vector-meson couplings to the octet baryons are enhanced in SU(3) symmetry, 
the extension from SU(6) to SU(3) symmetry hardens the EoS very much.  
In each model, the hardest EoS is given by the case (3), while the softest one is obtained in the case (1).  
This tendency can be related to the fact that, as seen in Figs.~\ref{fig:Composition-QMC} - \ref{fig:Composition-FSU}, 
the densities at which the hyperons appear in the case (3) are rather higher than those in the case (1).  
In general, the strange mesons, especially the $\phi$ meson, also play an important role in supporting a heavy neutron star.  
In the mass-radius relations presented in Figs.~\ref{fig:TOV-QMC-CQMC} - \ref{fig:TOV-FSU-IU-FSU}, 
we can again see that, in each model, the maximum neutron-star mass in the case (3) is heaviest, while the lightest one is given in the case (1).  
We note that, in Fig.~\ref{fig:TOV-GM1-GM3}, the curve (red solid) for the NL3 model in SU(3) symmetry cannot reach the maximum point 
because the nucleon mass becomes negative at the density before the maximum point.  

We summarize the following, several comments on the mass-radius relations shown in Figs.~\ref{fig:TOV-QMC-CQMC} - \ref{fig:TOV-FSU-IU-FSU}.  
In the QMC and CQMC models, the maximum neutron-star masses calculated in SU(3) symmetry are consistent with the pulsar PSR J1614-2230.  
In particular, the mass in the CQMC model clearly exceeds the mass of PSR J1614-2230.  
Because the difference between the QMC and CQMC models is originated by the hyperfine interaction between two quarks inside a baryon, 
the large difference between the two maximum masses is mainly generated by this microscopic interaction.  
It is noticeable that the quark-quark hyperfine interaction is very vital to obtain the correct mass spectra of octet baryons 
in a nuclear medium~\cite{Nagai:2008ai,Miyatsu:2010zz}.  

In the GM1 model, the maximum neutron-star mass in SU(3) symmetry is much larger than the observed mass of PSR J1614-2230.  
In contrast, the maximum mass in the GM3 model is clearly under the observed value (see the left panel in Fig.~\ref{fig:TOV-GM1-GM3}).  
The difference between the two models is just in the values of the nuclear incompressibility and the slope parameter, 
namely $K_v = 300 \, (240)$ MeV and $L = 93.9 \, (89.7)$ MeV for the GM1(3) model (see Table~\ref{tab:cc-non-linear-sigma}).  

The NL3 model is a unique model, 
in which the mass of PSR J1614-2230 can be explained even in SU(6) symmetry (see the right panel in Fig.~\ref{fig:TOV-GM1-GM3}).  
This model may be characterized by the large values of symmetry energy ($a_4 = 37.4$ MeV) and slope parameter ($L = 118$ MeV) (see Table~\ref{tab:cc-non-linear-sigma}).  

The rather large values of $a_4$ and $L$ are also used in the TM1 model, where only the SU(3) result can, however, reach the mass of PSR J1614-2230.  
Furthermore, the difference between the maximum masses in SU(6) and SU(3) symmetries is very large in the TM1 model (see also Table~\ref{tab:NS-properties}).  
This fact may be caused by the repulsive force due to the nonlinear $c_{3} {\bar \omega}^4$ term in Eq.(\ref{eq:Lagrangian-NL}).  
Note that, to reproduce the same saturation condition as in SU(6) symmetry, 
the strength of $c_{3}$ in SU(3) symmetry is larger than that in SU(6) symmetry (see Table~\ref{tab:cc-non-linear-sigma-omega-rho}).  

Unfortunately, the mass of PSR J1614-2230 cannot be explained by the FSUGold and IU-FSU models.  
However, in both models the maximum mass in SU(3) symmetry becomes $1.8 - 1.9 M_\odot$, which is not far from the observed mass.  
In these models, the maximum mass in SU(6) symmetry is again very different from the value in SU(3) symmetry (see Table~\ref{tab:NS-properties}).  
Furthermore, although the curves for $M/M_\odot$ in the SU(6) and SU(3) cases normally coincide with each other 
in the low mass region (see Figs.~\ref{fig:TOV-QMC-CQMC} - \ref{fig:TOV-GM1-GM3}), 
the two curves clearly stay away from each other even at $M/M_\odot = 0.8$ in the FSUGold and IU-FSU models (see Fig.~\ref{fig:TOV-FSU-IU-FSU}).  
These facts may again be caused by the very large difference between the values of $c_{3}$ in SU(6) and SU(3) symmetries (see Table~\ref{tab:cc-non-linear-sigma-omega-rho}).  

\section{Summary}
\label{sec:summary}

We have calculated the particle fractions, the meson fields and the EoS in the core of a neutron star, 
using the popular RMF models (such as GM1, GM3, NL3, TM1, FSUGold and IU-FSU) as well as the QMC and CQMC models.  
It is noticeable that some of the RMF models are accurately parameterized to compute the properties of infinite nuclear matter and finite nuclei.  
On the other hand, because, in the QMC and CQMC models, the quark degrees of freedom in a baryon are taken into account, 
they allow us to consider the variation of the quark structure of baryon in dense mater.  
In particular, the CQMC model involves the quark-quark hyperfine interaction, and thus it can correctly describe the octet baryon spectra in matter as well as in vacuum.  

In the present calculations, we have examined the extension from SU(6) spin-flavor symmetry based on the quark model 
to SU(3) flavor symmetry in determining the isoscalar, vector-meson couplings to the octet baryons.  
We have also studied how the strange mesons ($\sigma^{\ast}$ and $\phi$) contribute to the internal structure of a neutron star.  

In SU(3) symmetry, we have found that the models except GM3, FSUGold and IU-FSU can explain the mass of PSR J1614-2230.  
In the GM3, FSUGold and IU-FSU models, although the maximum mass cannot reach $1.97\pm0.04 M_{\odot}$, the calculated mass is not far from that value.  
Therefore, the extension from SU(6) to SU(3) symmetry and the strange vector meson, $\phi$, are very significant in sustaining a heavy neutron star.  
In addition, the variation of baryon structure in matter also helps prevent the collapse of a neutron star.  

In RMF models, the NL potential, $U_{NL}$, is indispensable for reproducing the saturation condition for symmetric nuclear matter and the properties of finite nuclei.  
In the present calculations, it involves not only the usual, nonlinear $\sigma$ terms 
but also the $c_{3} {\bar \omega}^4$ term and the isoscalar-isovector $\Lambda_{\omega \rho} {\bar \omega}^2 {\bar \rho}^{2}$ term.  
Among them, in particular, the nonlinear $c_{3} {\bar \omega}^4$ term hardens the EoS, and thus enhances a neutron-star mass.  
Furthermore, the nonlinear isoscalar-isovector coupling plays a unique role in the particle fractions inside a neutron star.  
Because the $\sigma$-$\Sigma$ and $\sigma^{\ast}$-$\Sigma$ coupling constants are usually determined 
so as to fit the (repulsive) mean-field potential for the $\Sigma$ in nuclear matter, 
it becomes difficult to create the $\Sigma$ hyperon in the core of a neutron star.  
However, in the FSUGold and IU-FSU models, instead of the $\Xi^{0,-}$, the $\Sigma^{-}$ can emerge with a considerable fraction even at rather low density.  
As the power counting~\cite{Mueller:1996pm,Furnstahl:1997} suggests, 
there may be many possible, NL couplings and many-body forces containing various meson fields, which may contribute to the EoS.  
It is thus interesting to study how such interactions contribute to the properties of a neutron star.  We note that, at the MF level, the NL potential may be 
regarded as many-body interactions among baryons because the meson fields are just auxiliary fields and thus they can be replaced with bilinear forms of baryon fields. 

In RMF models, the parameterization is usually performed using experimental data measured around $n_B^0$.  
However, because the region of density which is important in the EoS for a neutron star may be the region of $0.8 - 1.1$ fm$^{-3}$ ($> 6 n_B^0$), 
no one knows if such parameterizations work correctly at such high densities.  
Therefore, although in this paper we have studied RMF models from various standpoints, it may be difficult to winnow the good models out at the RMF level.  
As suggested in Dirac-Brueckner-Hartree-Fock calculations~\cite{vanDalen:2011ze}, 
it may, at least, be imperative to include the density dependence of the parameters to obtain conclusive results on the EoS.  

In the present calculations, we have not considered the Fock (exchange) term.  
Although, in {\it naive} QHD-I~\cite{Serot:1984ey}, the Fock contribution seems very small in symmetric nuclear matter, 
it plays a very important role even around the normal nuclear matter density as well as at high densities, 
if the $\rho$ meson is included and the tensor interaction thus arises~\cite{Miyatsu:2011bc,Katayama:2012ge}.  
It is remarkable that, when the tensor interaction is taken into account, 
the $\Lambda$ hyperon does not appear in the core even at high density~\cite{Miyatsu:2011bc,Katayama:2012ge,Whittenbury:2012rn}.  
Furthermore, the tensor contribution is very important in reproducing the density dependence of symmetry energy, $a_4$.  

At very high density, the quark and gluon degrees of freedom, rather than the hadron degrees of freedom, may take place in the core matter~\cite{Glendenning:2000}.  
Because the degrees of freedom in quark-gluon matter are generally large, 
it is necessary to assume a rather strong correlation between quarks and gluons to support a massive neutron-star mass.  
It would be very interesting to investigate how the quark-gluon phase connects with the hadron one and how such degrees of freedom contribute to the EoS for a neutron star.  

\begin{acknowledgments}
This work was supported by the National Research Foundation of Korea (Grant No. 2011-0015467).
\end{acknowledgments}



\end{document}